\documentclass[mres, copyrightpage]{mqthesis}

\usepackage{breqn}
\usepackage{graphicx}
\usepackage{float}
\usepackage{color}

\usepackage{epstopdf,cancel}
\usepackage{epsf,latexsym,bbm,euscript}
\usepackage{amssymb,amsmath}
\usepackage{mathrsfs}
\usepackage{bm}
\usepackage{mathtools} 
\usepackage{times,graphics}
\usepackage{caption}
\usepackage{subcaption}
\usepackage{romannum}

\usepackage{mathtools}
\usepackage{url,hyperref}
\newcommand{\etal}{\textit{et al.}}

\begin{document}

\frontmatter

\title{Properties of space-time in the vicinity of trapped regions}
\author{Pravin Kumar Dahal}
\department{Physics and Astronomy}  

\titlepage

\chapter{Acknowledgements}

Firstly, I would like to thank my supervisor A/Prof. Daniel Terno, for all his help in making this thesis possible. I would then,  like to thank Prof. Joanne Dawson and the administrative staff of Macquarie University for their help in carrying out this research project. I would also like to thank Sebastian Murk for both his scientific and general assistance and AJ Terno for proofreading this manuscript.

I am supported by the IMQRES scholarship of Macquarie University to carry out this research project.

I finally would like to thank all of my family, especially my wife Elija Timalsina, for all their moral and social support and encouragement. This path in the pursuit of passion would have been impossible without their support.

\let\cleardoublepage\clearpage
\chapter{Abstract}

We study the near-horizon geometry in axisymmetric space-times. The general axisymmetric metrics have seven different parameters depending on three coordinates, and consistency analysis of the Einstein equations on such space-times is a formidable task. After discussing some general aspects of axially symmetric metrics, we begin the investigations of the simplest representative of this class: the Kerr–Vaidya metrics. Kerr–Vaidya metrics can be derived from the Vaidya metric by the complex coordinate transformation suggested by Newman and Janis. We show that the energy-momentum tensor belongs to type \Romannum{3} in the Segre-Hawking-Ellis classification but has a special form with all Lorentz-invariant eigenvalues belonging to zero. It is known that the apparent horizon of the outgoing Kerr–Vaidya metric coincides with that of the Kerr space-time and that it is not so for the ingoing metric. We find its location for quasi-stationary Kerr–Vaidya black holes. The energy-momentum tensor of the Kerr-Vaidya geometries violates the null energy condition. We show that similar to the spherically symmetric accreting black hole, energy density, pressure, and flux for an infalling observer are diverging in the outgoing Kerr–Vaidya metric. This firewall leads to the violation of a specific quantum energy inequality that bounds the violation of the null energy condition.

\let\cleardoublepage\clearpage

\tableofcontents


\mainmatter

\begin{savequote}[45mm]
You have a right to perform your prescribed duties, but you are not entitled to the fruits of your actions. Never consider yourself to be the cause of the results of your activities, nor be attached to inaction. 
\qauthor{Gita: Chapter 2, Verse 47}
\end{savequote}

\chapter{Introduction}

There are billions of massive dark compact objects existing in our Universe. About a hundred of them are astrophysical black hole candidates with precisely known values of mass and spin~\cite{b2}. The question, however, is if they possess the distinctive black hole features as envisaged in classical general relativity~\cite{43}. 

\section{Classical black holes}

The idea of a black hole dates back to 1783 when John Michell considered stars with such a strong gravitational pull that even light could not escape. This idea, however, remained mostly as a curiosity until Einstein formulated the general theory of relativity. Black holes of general relativity are among the simplest solutions of the Einstein equations and one of its most profound predictions. They form the foundation of our ideas about astrophysical ultra-compact objects. One of their remarkable features is an event horizon, which is a boundary separating the interior of the black hole that is inaccessible from the external/outside Universe. The events occurring beyond the event horizon are causally disconnected from the outside world. Classical black holes, as shown by Penrose and Hawking, contain space-time singularities~\cite{3}. In a globally hyperbolic space-time, according to the Penrose singularity theorem~\cite{stp65}, the existence of a future trapped surface and the null energy condition implies the existence of a space-time singularity. Singularity is the region of space-time where classical general relativity predicts the infinite value of the gravitational field. While there are several not fully equivalent definitions of singularity, one way to characterize it invariantly is by diverging values of the curvature scalars. Singularity represents a breakdown of classical general relativity. A cosmic censorship conjecture of Penrose~\cite{4} posits that singularities should be enclosed inside the event horizon. So, the censorship conjecture implies that the formation of the singularity does not influence the outside observer. The Schwarzschild and the Kerr solutions are the two best-known examples of classical black hole solutions of empty space-time, which we will discuss in Sec.~\ref{sk1.3}. From a classical perspective, they are an asymptotic state of spherically symmetric and axisymmetric gravitational collapse of objects greater than three times the mass of the sun~\cite{6}.

Despite its conceptual simplicity, an event horizon is an undetectable feature of a black hole by definition~\cite{ah4}, as we need to know the entire history of space-time to locate it, and this is impossible for any mortal observer. Because of its teleological nature~\cite{ah2,ah3}, we may be crossing the event horizon here and now, but we would not know it, thereby necessitating an operationally meaningful feature for observation and testing. An apparent horizon, a boundary enclosing the trapped region from which currently not even light can escape, provides such a workable definition for numerical calculations. The quasi-local nature of the apparent horizon~\cite{ah10} makes it a suitable candidate characterizing a physical black hole~\cite{pbh8}, which is a domain of space-time where outward-pointing null geodesics bend backward. It may or may not have the singularity constituting a regular black hole or the event horizon. We give a detailed introduction of both the event and apparent horizons in Sec.~\ref{h11}.

\section{Quantum effects}

In the absence of a fully developed quantum theory of gravity, the analysis of black holes is largely performed using semiclassical physics (Sec.~\ref{st14}). The semiclassical approach was pioneered by the likes of Hawking, Parker, Fulling, Davies, and Dewitt in the late 1960s, and its most celebrated prediction is the Hawking radiation~\cite{5}. Close to a collapsing mass, the quantum field in the vacuum state in the asymptotic past need not be in a vacuum state near the future. A distant observer, at a later time, will see a flux of radiation emitted from the collapsing body known as Hawking radiation~\cite{7}. The mass of the collapsing body decreases because of the radiation outflux, and this entails taking an additional semiclassical backreaction term for solving the Einstein equation~\cite{26,sg13}. As long as the radiation outflux is small enough, the backreaction can be modeled self-consistently by the vacuum expectation value of the stress-energy tensor~\cite{sg60}. There are numerous ways to derive Hawking radiation, all of which are based on three basic assumptions~\cite{sg13,ilp18}: 1) Einstein classical equation of general relativity describes the gravitational field, 2) the backreaction of the outgoing radiation on the space-time geometry is small enough such that it can be treated as a perturbative term, and 3) the field of the emitted radiation is distinct from that of the collapsing matter. The approximately thermal nature of the radiation thus emitted is related to the fact that it experiences exponential redshift in traveling from near the event horizon to infinity. Hawking radiation provided the foundation for the development of black hole thermodynamics and thus was the first result to show the connection between gravity, quantum field theory, and thermodynamics.

There are, however, some intricacies in introducing quantum fields to the total stress-energy tensor of the Einstein equation. Hawking~\cite{5} showed that particle production is impossible unless the dominant energy condition (Sec.~\ref{e11}) is violated. Quantum field theory in curved space-time allows space-time curvature to induce the negative value of energy, momentum, and stress, which can violate the energy conditions. As quantum states can violate the energy conditions, quantum effects allow the violation of both the Penrose-Hawking singularity theorem and Penrose's cosmic censorship conjecture~\cite{8}. The reason behind this is that the classical laws of black hole mechanics and the singularity theorems are based on the assumption that at least the null energy condition is satisfied. Consequences include the existence of a trapped region without a singularity, as well as the existence of a naked singularity (singularity not enclosed by an event horizon). The violation of energy conditions is, however, bounded by the quantum energy inequalities, which are defined in Sec.~\ref{113}. Besides, the emission of Hawking radiation by the violation of the dominant~\cite{5} and hence the null energy condition allows a decrease in the black hole's area, which necessitated the generalization of the area theorem. The original area theorem states that the surface area of a black hole can not decrease~\cite{ah5}. According to Hawking and Ellis~\cite{10}, the violation of the null energy condition makes an apparent horizon accessible to a distant observer.

Hawking radiation is also the root of the information loss problem~\cite{ilp59,ilp17}, which is an apparent contradiction between quantum mechanics and general relativity. This disagreement is an apparent consistency paradox: quantum field theory is built as a unitary theory, and the emission of nearly thermal radiation from the relatively ordered collapsing matter causes loss of information, which is interpreted as a violation of the unitarity. Among numerous resolutions of this problem, one is the possible occurrence of the firewall where an infalling observer encounters high energy quanta at the horizon~\cite{ilp17, ilp19}. Furthermore, the emission of Hawking radiation does not require an event or even an apparent horizon, but only an ongoing collapse. So, any compact horizonless objects can emit Hawkwing radiation~\cite{9}, thereby adding more complexity to the classical picture of a black hole. If this emission process continues even after the size of a collapsing object reaches the Planck scale, the realization of black holes as an asymptotic state of gravitational collapse becomes unattainable (see, for example, Ref.~\cite{sbh60}). Numerous views allowing different scenarios for the final stage of the collapse exist in literature~\cite{ilp17,ilp19}, and we are not dealing with this problem here.

\section{Alternatives to black holes}

The interior of the black hole contains a space-time singularity enclosed by the horizon, because of which the future evolution of a well-defined initial state becomes uncertain. Thus, considering the possibility of the different endstate of collapse might be the resolution of these problems, which naturally directs us in search of alternatives to black holes.

\begin{figure}[!htbp]
\centering
\includegraphics[width=0.95\textwidth]{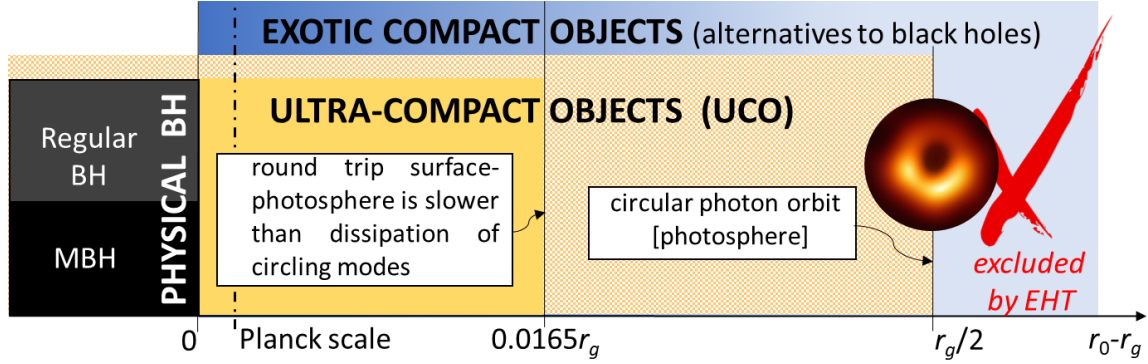}
\caption{Possible candidates for ultra-compact objects categorized based on the dynamical features that depend on their size relative to the Schwarzschild radius $r_g$. MBH represents mathematical black holes. (After Ref.~\cite{42})}
\label{uco1}
\end{figure}

Alternatives of black holes are various exotic compact objects~\cite{42, eco9}, which are the horizonless objects resulting from the gravitational collapse. Their explicit models often involve strongly non-classical matter and deviations from known physics. Fig.~\ref{uco1}
shows the schematic representation of the collection of ultra-compact objects whose dynamical properties are expressed in terms of their radius with respect to the Schwarzschild radius $r_g$. Like black holes, they tend to be dark, and the high frequency electromagnetic or gravitational wave could have a circular orbit around them. This circular orbit characterizes a surface called photosphere, which is located at $r-r_g=r_g/2$. Exotic compact objects with the photosphere are termed as ultra-compact objects. The photosphere defines the so-called ``shadow" of an object, which is basically how an object looks like when illuminated by the accretion disk. The photosphere also determines the geodesic description of the high frequency gravitational and electromagnetic waves near the horizon. Light is infinitely red-shifted in the coordinate frame and thus never reaches the horizon of the black hole from the photosphere. Exact same geodesic description from an ultra-compact object with $r-r_g<0.0165 r_g$ is expected around the photosphere.

So, the classification of different classes of ultra-compact objects depends on the behavior of the light field around their photosphere. Although the existence of such objects has been verified from both the electromagnetic and gravitational observations~\cite{eo10, go11}, whether they are physical black holes or exotic compact objects are not yet confirmed.

\section{Semiclassical theory}\label{st14}

Far from a Planck scale, in the absence of the satisfactory quantum theory of gravity, we assume the validity of a self-consistent semiclassical approach that assumes the existence of quantum fields on the classical space-time geometry. In this approach, the dynamics of the gravitating system are described by the Einstein equation, where the Einstein tensor is equal to the expectation value of the renormalized stress-energy tensor~\cite{sg13}. The renormalized stress-energy tensor is the averaged sum total of all the matter fields and the quantum excitations. So, the following implicit assumptions of the semiclassical framework~\cite{sf61,sf62} are necessary for our project:
\begin{itemize}
    \item The classical notion of the gravitational field, which is described by the metric $g_{\mu\nu}$ and its derivatives is correct. So, classical concepts like trajectory and singularity are used
    \item Quantum effects should be accounted for to describe the gravitational field. This entails us to write the Einstein field equation
    \begin{equation}
        G_{\mu\nu}=8\pi \left<T_{\mu\nu}\right>,
    \end{equation}
    where $G_{\mu\nu}$ is the Einstein tensor and $\left<T_{\mu\nu}\right>$ is the total expectation value of the energy-momentum tensor representing both the collapsing matter and the quantum excitations.
\end{itemize}

In this project, we aim to clarify some issues in the self-consistent approach. For this, we first assume that the black hole exists (in the broadest possible sense). The existence of the black hole is physically relevant if it is formed in a finite time. Second, we assume that the boundary of the black hole is regular, an assumption which is valid in classical general relativity. The regularity of the apparent horizon is characterized by the curvature invariants being finite there. The results were confined to the case of spherical symmetry (see Sec.~\ref{s3}). We aim to generalize these results to the case of axial symmetry, starting from the simplest models of the Kerr-Vaidya metrics.

\section{Trapped surfaces}

Trapped surfaces are the quasi-local entities that are used to characterize a black hole. A detailed overview to calculate the expansion of null curves is given in Ref.~\cite{ah5}.

\subsection{Expansion of null congruence}

Deformation of a two-dimensional medium serves as a template for general relativistic analysis. Consider a small displacement $\xi^\alpha$ about some reference point $O$. The rate of deformation corresponding to  $\xi^\alpha$ is linear and can be written as
\begin{equation}
    \frac{d\xi^\alpha}{d t}=B^\alpha_\beta(t) \xi^\beta,
    \label{d2}
\end{equation}
where $B^\alpha_\beta(t)$ is the tensor describing the deformation and the displacement in a small time interval $\Delta t=t_1-t_0$ becomes
\begin{equation}
    \xi^\alpha(t_1)=\xi^\alpha(t_0)+B^\alpha_\beta(t_0) \xi^\beta(t_0)\Delta t .
    \label{d3}
\end{equation}
In general, the tensor $B^\alpha_\beta$ can be written as
\begin{equation}
    B^\alpha_\beta=\frac{1}{2}\vartheta\delta^\alpha_\beta+ \sigma^\alpha_\beta+\omega^\alpha_\beta,
\end{equation}
where the expansion scalar $\vartheta=B^\alpha_\alpha$ is the trace of $B^\alpha_\beta$, the shear tensor $\sigma_{\alpha\beta}=B_{\alpha\beta}+B_{\beta\alpha}-\frac{1}{2}\vartheta\delta_{\alpha\beta}$ is the symmetric tracefree part of $B^\alpha_\beta$ and the rotation tensor $\omega_{\alpha\beta}=B_{\alpha\beta}-B_{\beta\alpha}$ is the antisymmetric part of $B^\alpha_\beta$. The only deformation we want to consider here is expansion, and for this we assume $B^\alpha_\beta$ proportional to the identity matrix
\begin{equation}
    B^\alpha_\beta=\begin{pmatrix}
\frac{1}{2}\vartheta & 0 \\
0 & \frac{1}{2}\vartheta
\end{pmatrix}.
\label{d4}
\end{equation}
If we treat Eq.~\ref{d3} as the translation of coordinates from $\xi^\alpha(t_0)$ to $\xi^\alpha(t_1)$, then the Jacobian of this transformation is given as
\begin{equation}
    J=\det[\delta^\alpha_\beta+B^\alpha_\beta \Delta t]= 1+\vartheta \Delta t.
\end{equation}
Thus, the area of the surface at time $t_1$ due to the deformation becomes
\begin{equation}
    A(t_1)=(1+\vartheta \Delta t) A(t_0),
\end{equation}
which gives:
\begin{equation}
    \vartheta=\frac{1}{A(t_0)}\frac{A(t_1)-A(t_0)}{\Delta t}.
\end{equation}
Thus, $\vartheta$ gives the fractional area change, which can be viewed as the expansion of the surface.

\subsubsection{Congruence of null geodesics}

A curve in the casually connected region is said to be future-directed if it can be joined from a point to any other point in the future. A past directed curve is likewise defined. In this work, we consider only orientable space-times~\cite{10}.

A congruence in a region of space ${\cal O}$ is defined as the set of integral curves with non-vanishing vector field at each point. Let the component of the vector field tangential to the surface be $l^\alpha$ and let $\xi^\alpha$ be the deviation vector that is orthogonal to $l^\alpha$ and Lie transported along it at the surface. As our goal is to calculate the expansion of the field $l^\alpha$, we first need to isolate the purely transverse part of the vector $\xi^\alpha$. This is because $l^\alpha \xi_\alpha=0$ is not a sufficient condition for the vector to have component purely transverse to $l^\alpha$. We thus choose another null vector field $n^\alpha$ such that $l^\alpha n_\alpha=-1$. It allows us to write the transverse metric as
\begin{equation}
    h_{\alpha\beta}=g_{\alpha\beta}+l_\alpha n_\beta + n_\alpha l_\beta,
    \label{d8}
\end{equation}
where
\begin{equation}
    h_{\alpha\beta}l^\alpha=h_{\alpha\beta}n^\beta=0, \qquad h_{\alpha}^\alpha=2, \qquad h^\mu_\alpha h^\beta_\mu=h^\beta_\alpha,
\end{equation}
thereby showing that $h_{\alpha\beta}$ is a metric orthogonal to both $l^\alpha$ and $n^\alpha$ and is two dimensional.

As the deviation vector is Lie transported along the null vector field, its Lie derivative along $l^\alpha$ is zero:
\begin{equation}
    \xi^\alpha_{;\beta}l^\beta=l^\alpha_{;\beta}\xi^\beta,
     \label{d10}
\end{equation}
where semicolon ($;$) denotes the covariant derivative. Thus, if we define $B^\alpha_\beta=l^\alpha_{;\beta}$, then from Eq.~\eqref{d10}, we can see that $B^\alpha_\beta$ measures the change  of deviation vector along the direction of the vector field. $l^{\alpha}$ being tangent vector to the null geodesic at some surface satisfies the geodesic equation on that surface giving $l^\beta l^\alpha_{;\beta}=0$. This shows that $B^\alpha_\beta$ is orthogonal to $l^\alpha$. However, $B^\alpha_\beta$ is not orthogonal to $n^\alpha$ in general. Thus, $B^\alpha_\beta$ also has some non-transverse components which should be removed. This can be achieved by using the transverse metric $h_{\alpha\beta}$,
\begin{equation}
    \tilde{B}_{\alpha\beta}\vcentcolon=h^\mu_\alpha h^\nu_\beta B_{\mu\nu},
    \label{d11}
\end{equation}
where $\tilde{B}^\alpha_\beta$ has only transverse components. $\tilde{B}^\alpha_\beta$ defined here is equivalent to the one introduced in Eq.~\eqref{d2} as the measure of the deformation. As we are only considering the case of expansion, we want the trace of $\tilde{B}^\alpha_\beta$ to give $\vartheta$ as in Eq.~\eqref{d4} and for this
\begin{equation}
    \tilde{B}_{\alpha\beta}=\frac{1}{2}\vartheta h_{\alpha\beta},
    \label{d12}
\end{equation}
thereby giving $\vartheta=h^{\alpha\beta}\tilde{B}_{\alpha\beta}=h^{\alpha\beta}B_{\alpha\beta}$ as the expansion of the null tangent vector on the surface where it satisfies the geodesic equation.

We thus have the two equivalent relations to calculate the expansion of null geodesic congruence on a two dimensional surface~\cite{ah5}:
\begin{itemize}
    \item The first is the ab-initio approach which is by the direct calculation of the congruence cross-sectional area
    \begin{equation}
        \vartheta=\frac{1}{A(\lambda)}\lim_{\Delta\lambda\to0}\frac{\Delta A(\lambda)}{\Delta \lambda} \label{exp114},
    \end{equation}
    where $A(\lambda)$ is purely transversal and $\lambda$ is some parameter depending on initial time $t_0$.
    \item The second is by the calculation of the transverse metric $h_{\alpha \beta}$ and the deviation vector $B^\alpha_\beta=l^\alpha_{;\beta}$
    \begin{equation}
        \vartheta=h^{\alpha\beta}l_{\alpha;\beta} \label{exp115}.
    \end{equation}
\end{itemize}


\subsection{Event and apparent horizons}\label{h11}

Asymptotically flat space-time containing classical black holes can be separated into two regions, the region from which all null curves can reach the future null infinity and the region from which no null curve reaches the future null infinity. A boundary separating these two regions of space-time is defined as an event horizon. This is a null hypersurface. To locate an event horizon, one should know the complete information of the future null infinity and its global past, making it unobservable~\cite{ah4}. This global nature of an event horizon also makes it a less useful entity for the study of dynamical space-times. A quasi-local notion of a black hole that at least in principle can be verified by a finite number of spatially and temporally localized observations is useful both conceptually and practically. It also captures the original black hole idea.

We take a future-directed field of ingoing and outgoing null vectors $l^\alpha$ and $n^\alpha$ respectively, that are orthogonal to the two-dimensional spacelike surface (with spherical topology) satisfying $l^\alpha n_\alpha=-1$. We can calculate the corresponding expansions $\vartheta_l$ and $\vartheta_n$ associated with these null vectors, among which the condition $\vartheta_l=0$ is pivotal for the definition of the trapped surfaces. A surface is said to be trapped if both $\vartheta_l<0$ and $\vartheta_n<0$. Thus on the trapped surface, both the ingoing and outgoing null geodesics are converging. A marginally outer trapped surface (MOTS) has $\vartheta_l=0$ and $\vartheta_n<0$. This surface outlines the boundary from which the outgoing null geodesics starts to converge. The three-dimensional surface which can be foliated entirely by the MOTS is called a marginally outer trapped tube (MOTT). This MOTT is defined as the future apparent horizon in recent literature~\cite{ah6,ah7} and we will adopt this definition here. Other useful definitions of quasi-local horizons can be found in the literature. The non-exhaustive list of horizons are given in Refs.~\cite{ah2,ah64}. In Sec.~\ref{sk1.3}, we provide some simple examples of the trapped surfaces. Note that the examples given there are straightforward as the null vector fields are defined/guessed for the bulk of the space-time and not just on some hypersurface.

The definitions of MOTT provide an easy way of identifying black holes, and these are just by the calculation of null expansions $\vartheta_l$ and $\vartheta_n$. We thus do not have to wait till the end of time to know the formation of an event horizon to identify black holes. Trapped surfaces indeed turned out to be convenient and unique (for spherically symmetric foliation, which is an obvious choice of foliation because of the symmetry of space-time) for the numerical and analytical study of spherically symmetric space-times. However, the complication starts to appear when we consider foliations that are not spherically symmetric, even to calculate the trapped surface in spherically symmetric space-times. The foliation dependence of trapped surface in Vaidya space-time has been demonstrated by numerical calculations with some special choice of non-spherically symmetric foliation in Refs.~\cite{ah8,ah9}. Discussion of the foliation dependence of the trapped surface in Schwarzschild and Vaidya space-time can also be found in the review by Krishnan~\cite{ah10}. We will describe the trapped surface in Kerr-Vaidya space-time on $v/u=\text{constant}$ foliation in Ch.~\ref{ah31}.


\section{Schwarzchild, Vaidya and Kerr black hole solutions}\label{sk1.3}

A year after Einstein formulated the general theory of relativity, Karl Schwarzschild provided a vacuum solution of Einstein field equation that, in modern terms, describes the space-time outside a stationary black hole~\cite{s1}. The Schwarzschild solution in spherical coordinates can be written as:
\begin{equation}
    ds^2=-\left(1-\frac{2 M}{r}\right)dt^2+ \frac{r}{r-2 M}dr^2+ r^2 d\theta^2+ r^2 \sin^2\theta d\phi^2,
    \label{s114}
\end{equation}
where $r$ is the areal radius, and $t$ is the geometrically preferred time defined by the divergence-free Kodama vector~\cite{kv73}. It is sometimes convenient to use the outgoing/ingoing Eddington-Finkelstein coordinates~\cite{33}
\begin{equation}
    du_{\pm}=dt\mp\frac{r}{r-2 M}dr,
\end{equation}
to write this metric in the form
\begin{equation}
    ds^2=-\left(1-\frac{2 M}{r}\right)du_{\pm}^2\mp 2 du_{\pm} dr+ r^2 d\theta^2+ r^2 \sin^2\theta d\phi^2.
    \label{os118}
\end{equation}
where the upper sign is for retarded ($u_+=u$) and the lower sign is for advanced ($u_-=v$) coordinates. This is the Schwarzschild metric written in retarded/advanced null coordinates (or outgoing/ingoing Eddington-Finkelstein coordinate). Now, let us allow the mass $M$ to be the function of retarded/advanced time $v/u$. We then obtain the Vaidya metric in retarded/advanced coordinates~\cite{35}
\begin{equation}
    ds^2=-\left(1-\frac{2 M(u_{\pm})}{r}\right)du_{\pm}^2\mp 2 du_{\pm} dr+ r^2 d\theta^2+ r^2 \sin^2\theta d\phi^2,
    \label{av117}
\end{equation}
and this is the radiating solution of the Einstein equation in spherical symmetry with a pure emitting/absorbing radiation field.

\subsection{Newman-Janis Transformation} \label{nj12}

Using Newman-Janis transformation~\cite{nj1} it is possible to obtain a more general axisymmetric vacuum solution of the Einstein equation from the spherically symmetric vacuum solution. For example, we can obtain the Kerr metric (Sec.~\ref{k162}) from the Schwarzschild metric and the Kerr-Newman metric from the Reissner-Nordstrom metric using this transformation. A brief derivation of the Kerr metric using the Newman-Janis transformation proceeds as follows.

The first step is to use the Schwarzschild metric in retarded coordinates as the so called seed metric. This metric can be written in terms of the null complex tetrad as
\begin{equation}
    g^{\mu\nu}=- l^\mu n^\nu- l^\nu n^\mu+ m^\mu \tilde m^\nu+ m^\nu \tilde m^\mu,
    \label{np119}
\end{equation}
where, two real and two complex null vectors are
\begin{equation}
    \begin{split}
        & l^\mu=\delta^\mu_r \qquad n^\mu=\delta^\mu_u- \frac{1}{2}\left(1-\frac{2 M}{r}\right) \delta^\mu_r\\
        & m^\mu=\frac{1}{\sqrt{2}r}\left(\delta^\mu_\theta+\frac{i}{\sin\theta}\delta^\mu_\phi \right) \qquad \tilde m^\mu=\frac{1}{\sqrt{2}r}\left(\delta^\mu_\theta-\frac{i}{\sin\theta}\delta^\mu_\phi \right),
        \label{nt121}
    \end{split}
\end{equation}
respectively. These four vectors satisfy the following completeness and orthogonality relations:
\begin{equation}
    \begin{split}
        &l^\mu l_\mu =l^\mu m_\mu =l^\mu \tilde m_\mu=0,\\
        &n^\mu n_\mu=n^\mu m_\mu =n^\mu \tilde m_\mu=0 = m^\mu m_\mu,\\
        &l^\mu n_\mu=-1=-m^\mu \tilde m_\mu.
        \label{nj125}
    \end{split}
\end{equation}
This null tetrad forms a starting point of the Newman-Penrose formalism~\cite{ch41}. The second step of the Newman-Janis algorithm is the complex coordinate transformation
\begin{equation}
    {x'}^\mu=x^\mu- i a \left(-\delta^\mu_r+\delta^\mu_u \right).
\end{equation}
Writing explicitly, this coordinate transformation is
\begin{equation}
    u'=u-i a \cos\theta, \qquad r'=r+i a \cos\theta, \qquad \theta'=\theta, \qquad \phi'=\phi,
\end{equation}
where the prime here denotes the new coordinate, not the derivative. As a result, the tetrad $Z^\mu=\left(l^\mu, n^\mu, m_\mu, \tilde m_\mu \right)$ transforms as ${Z'}^\mu_i = \left(\frac{\partial {x'}^\mu}{\partial x^\nu}\right) Z^\nu_i$ giving
\begin{equation}
    \begin{split}
        & {l'}^\mu=\delta^\mu_r \qquad {m'}^\mu=\frac{1}{\sqrt{2}(r+i a \cos\theta)}\left(i a \left(\delta^\mu_u- \delta^\mu_r \right)\sin\theta +\delta^\mu_\theta+ \frac{i}{\sin\theta} \delta^\mu_\phi \right)\\ & {n'}^\mu= \delta^\mu_u- \frac{\delta^\mu_r}{2}\left(1-\frac{2 M r}{\rho^2}\right) \qquad \tilde {m'}^\mu=\frac{1}{\sqrt{2}(r-i a \cos\theta)}\left(-i a \left(\delta^\mu_u- \delta^\mu_r \right)\sin\theta +\delta^\mu_\theta -\frac{i}{\sin\theta}\delta^\mu_\phi \right)
        \label{nt123}
    \end{split}
\end{equation}
where $\rho^2=r^2+a^2 \cos^2\theta$. Now, the metric in terms of this transformed tetrad is
\begin{equation}
    {g'}^{\mu\nu}=- {l'}^\mu {n'}^\nu- {l'}^\nu {n'}^\mu+ {m'}^\mu \tilde {m'}^\nu+ {m'}^\nu \tilde {m'}^\mu.
\end{equation}
The new line-element is axially symmetric
\begin{dmath}
ds^2=-\bigg(1-\frac{2 M r}{\rho^2}\bigg)du^2-2 du dr+\rho^2 d\theta^2-
    \frac{4 a M r \sin^2\theta}{\rho^2}d\phi du
    +2 a \sin^2\theta d\phi dr
    + \frac{(r^2+a^2)^2-a^2 \Delta \sin^2\theta}{\rho^2}\sin^2\theta d\phi^2,
    \label{ook127}
\end{dmath}
where $\Delta=r^2-2 M r+a^2$. This is the Kerr metric~\cite{2, tk2} in retarded coordinates.

\subsection{The Kerr space-time} \label{k162}

Here and in Sec.~\ref{ks163}, we review some essential features of the Kerr metric. The transformation
\begin{equation}
    du=dt-\frac{r^2+a^2}{\Delta}dr, \qquad
    d\phi \to d\phi+\frac{a}{\Delta}dr,
\end{equation}
brings the metric of Eq.~\eqref{ook127} to the Boyer-Lindquist form~\cite{32}
\begin{equation}
    ds^2=-\bigg(1-\frac{2 M r}{\rho^2}\bigg)dt^2-\frac{4 a M r \sin^2\theta}{\rho^2}dt d\phi
    +\frac{\rho^2}{\Delta}dr^2
    +\rho^2 d\theta^2+\frac{(r^2+a^2)^2-a^2 \Delta \sin^2\theta}{\rho^2}\sin^2\theta d\phi^2,
    \label{k40}
\end{equation}
where $\rho^2=r^2+a^2 \cos^2\theta$ and $\Delta=r^2-2 M r+a^2$. We again perform the following coordinate transformation to this metric, which is the generalization of the Eddington-Finkelstein advanced coordinates~\cite{33} in spherical symmetry
\begin{equation}
    dv=dt+\frac{r^2+a^2}{\Delta}dr,\qquad
    d\phi\to d\phi+ \frac{a}{\Delta}dr.
\end{equation}
This will give the line-element of the form
\begin{multline}
    ds^2=-\bigg(1-\frac{2 M r}{\rho^2}\bigg)dv^2+2 dv dr+\rho^2 d\theta^2-
    \frac{4 a M r \sin^2\theta}{\rho^2}d\phi dv
    -2 a \sin^2\theta d\phi dr+\\
    \frac{(r^2+a^2)^2-a^2 \Delta \sin^2\theta}{\rho^2}\sin^2\theta d\phi^2. \label{ki133}
\end{multline}
This is the Newman-Janis form~\cite{nj1} and is the first published form of the Kerr metric~\cite{2}.

\subsection{Examples of the event and the apparent horizons}\label{ks163}

    \noindent\textbf{Stationary Kerr and Schwarzschild space-times:}
    
    We take the following pair of null geodesics to calculate the apparent horizon for the Kerr metric given in Eq.~\eqref{k40}:
    \begin{equation}
        l_\mu=\frac{\Delta}{2\rho^2}\bigg(-1,\; \frac{\rho^2}{\Delta},\; 0,\; a \sin^2\theta \bigg),\qquad
        n_\mu=\bigg(-1,\; -\frac{\rho^2}{\Delta},\; 0,\; a \sin^2\theta \bigg).
    \end{equation}
    These null geodesics satisfy the following completeness and orthonormal relations
    \begin{equation}
    l_\mu l^\mu=n_\mu n^\mu=0, \qquad l_\mu n^\mu=-1.
    \label{oc3}
    \end{equation}
    Using Eq.~\eqref{d12} for the calculation of null expansions, we get
    \begin{equation}
        \vartheta_l=\frac{r \Delta}{\rho^4}, \qquad \vartheta_n=-\frac{2 r}{\rho^2}.
    \end{equation}
    From the definition, the apparent horizon is the solution of $\vartheta_l=0$, and this is $\Delta=0$. Since the event and the apparent horizon coincides for the stationary metric, this surface is also the event horizon. $\Delta=0$ is indeed a null surface can be seen from the fact that the vector $n^\mu$ normal to the surface $r=\text{constant}$ satisfies
    \begin{equation}
        g^{\mu\nu} n_\mu n_\nu=g^{r r}=\frac{\Delta}{\rho^2}.
    \end{equation}
    These results reduce to the one for the Schwarzschild space-time when $a=0$. Thus, $r=2 M$ is both the event and the apparent horizon of the Schwarzschild metric given in Eq.~\eqref{s114}.
    
    In the Kerr metric, there also exists a surface outside the event horizon where the Killing vector (infinitesimal generator of isometries) $\partial_t$ is null. This surface, called ergosurface, occurs at $\partial_t \partial_t=g_{t t}=0$, outside which the Killing vector $\partial_t$ is timelike and inside is spacelike. The region between the ergosurface and the event horizon is called the ergosphere. It is evident that for the Schwarzschild space-time, the ergosurface coincides with the event horizon.
    
    \noindent\textbf{Vaidya space-time:}
    
\begin{figure}[!htbp]
    \centering
    \begin{subfigure}{.49\textwidth}
    \centering
    \includegraphics[width=1.25\textwidth]{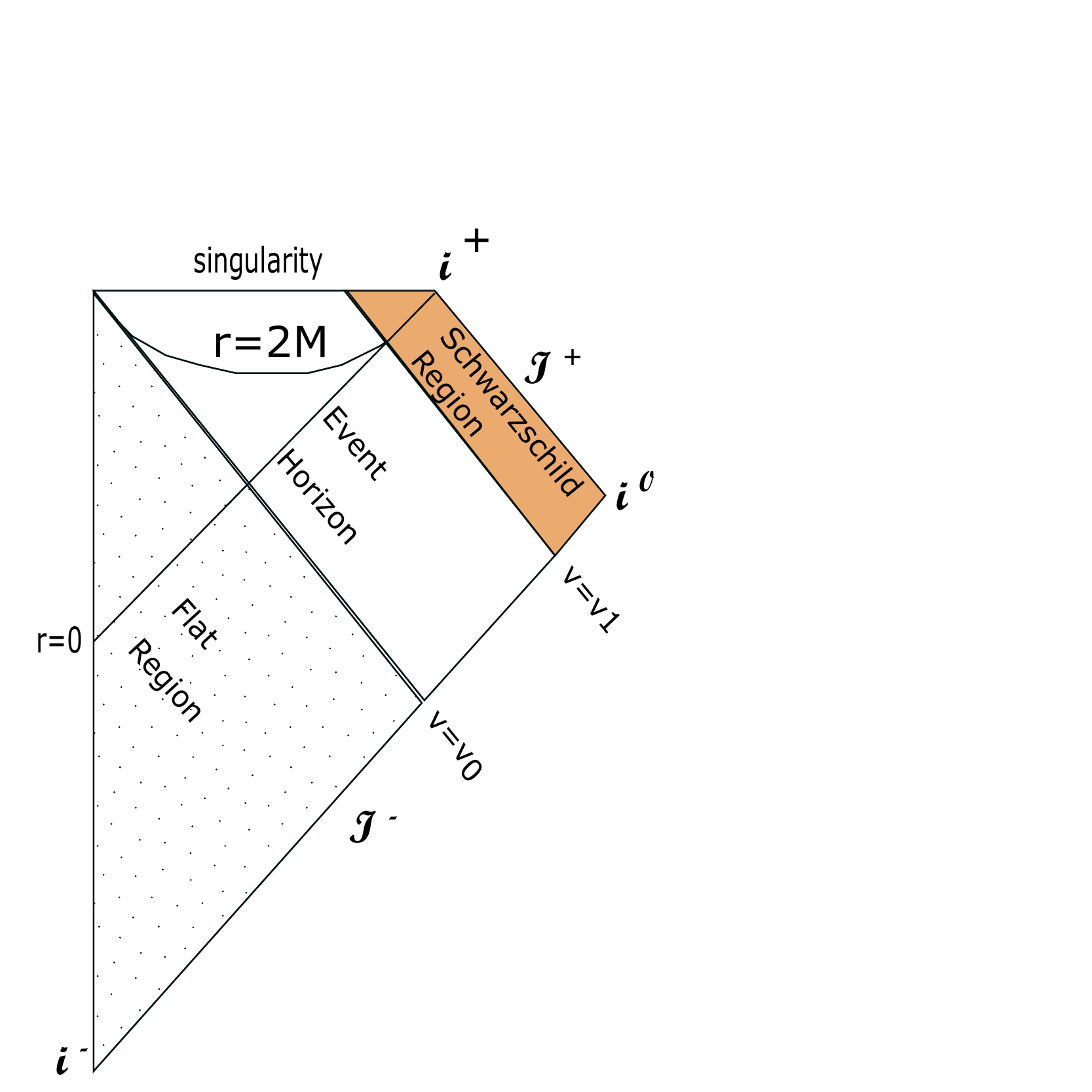}
    \caption{}
  \label{fig:sub1}
\end{subfigure}
\begin{subfigure}{.49\textwidth}
\centering
  \includegraphics[width=1.25\linewidth]{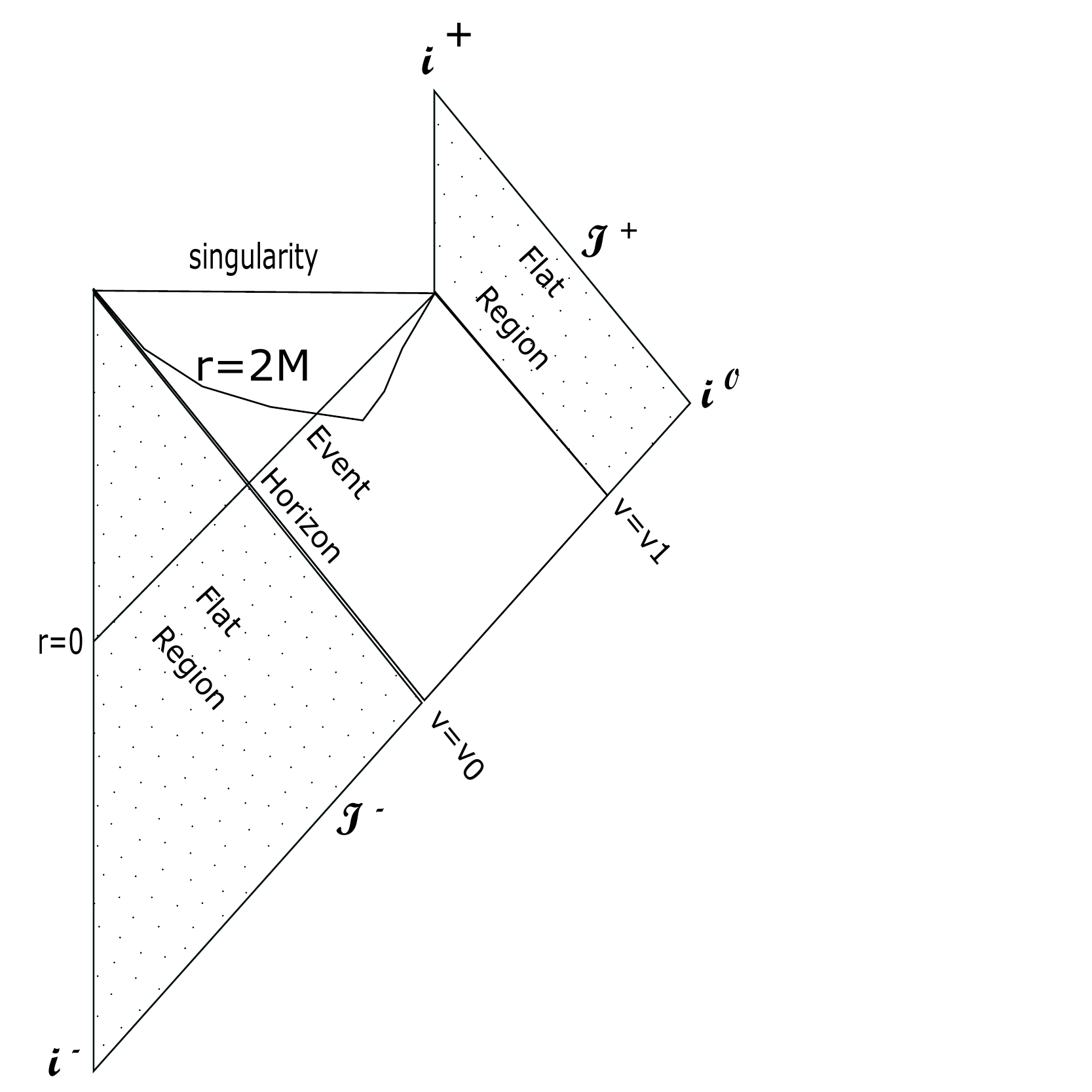}
  \caption{}
\end{subfigure}
    \caption{Penrose diagrams for spherically symmetric accreting null shell: a) Classical collapse of a null shell (the accretion onsets at $v_0$ and completes at $v_1$), and b) Accretion of an evaporating null shell (the black hole fully evaporates at $v_1$). Here, $\textit{i}^0$ denotes spacelike infinity, $\textit i^-$ past timelike infinity, $\textit i^+$ future timelike infinity, ${\cal J}^-$ past null infinity, and ${\cal J}^+$ future null infinity. The Minkowski region is dotted and the Schwarzschild region is shaded. $r=2 M(v)$ is the location of the apparent horizon for a general spherical symmetric black hole. Black hole evaporation violates the null energy condition, which is the necessary condition for the apparent horizon to lie outside the event horizon.}
    \label{pdv2}
    \end{figure}

    We take the following pair of null geodesics to calculate the apparent horizon for the Vaidya metrics given in Eq.~\eqref{av117}:
    \begin{equation}
        l_\mu=\frac{r-2M(o)}{2 r}\left(-1,\; \frac{r}{r-2 M(o)},\; 0,\; 0 \right), \qquad n_\mu=\left(-1,\; 0,\; 0,\; 0 \right),
    \end{equation}
    where $o=u, v$. These null geodesics satisfy the following completeness and orthonormal relations given in Eq.~\eqref{oc3}. Again, using Eq.~\eqref{d12} for the calculation of null expansions, we get
    \begin{equation}
        \vartheta_l=\frac{r-2 M(o)}{r^2}, \qquad \vartheta_n=-\frac{2}{r}.
    \end{equation}
    Clearly, the solution for $\vartheta_l=0$ is $r=2 M(o)$. So, $r=2 M(v)$ is the apparent horizon for the Vaidya metric in advanced coordinates and $r=2 M(u)$ is the apparent horizon for the Vaidya metric in retarded coordinates. Even for these simplest examples of the evolving space-times, locating the event horizon is not possible without the knowledge of the complete casual structure of the space-time. The Penrose diagram for the Vaidya metric in advanced coordinates depicting the relative position of the apparent and the event horizon (assuming that $M(v)$ is known, all the way, to the future) is shown in Fig.~\ref{pdv2}.
    The flat/Minkowski region in the space-time is because $M(v)=0$ and the Schwarzschild portion is because $M(v)=\mathrm{constant}$. So, the mass varies and the space-time evolves only in between.

\section{Energy conditions and the classification of the energy-momentum tensor}\label{e11}

Classical matter and fields have some `natural properties': they seem to obey certain constraints like the positivity of the energy density, show dominance of the energy density over the pressure, and that gravity is always attractive. These conditions are expressed mathematically in the form of the inequalities called energy conditions~\cite{10,ec20}. Energy conditions are critical for establishing many of the results of classical general relativity, while their violation by quantum systems leads to such as Casimir effect and Hawking's radiation. 

The occurrence of an apparent horizon implies the existence of an event horizon outside, provided that the null energy condition is satisfied~\cite{10}. Both of them coincide, only when space-time is stationary. For approximately stationary space-times, the locations of an event and apparent horizons would therefore be very close. The violation of the null energy condition points out to the possibility of the existence of the apparent horizon outside the event horizon, or even of the naked singularity. Here, we describe only some of the energy conditions which are relevant to our project, although there are more.

The \textit{weak energy condition} states that $T_{\mu\nu} u^\mu u^\nu \geqslant 0$ for any timelike $u^\mu$. This inequality captures the classical idea of the positivity of energy density.

The \textit{strong energy condition} states that $R_{\mu\nu} u^\mu u^\nu \geqslant 0$, for any timelike $u^\mu$. This condition ensures that gravity always attracts. This can be seen by using the Raychaudhari equation~\cite{re21} for a congruence of timelike geodesic
\begin{equation}
    \frac{d\vartheta}{d\tau}= -\frac{1}{3}\vartheta^2- \sigma^{\mu \nu} \sigma_{\mu \nu}+ \omega^{\mu\nu}\omega_{\mu\nu}-R_{\mu\nu}u^\mu u^\nu,
\end{equation}
where $\vartheta$ is the expansion scalar which is explained in details below, $\sigma^{\mu \nu}$ is the shear tensor, $\omega^{\mu\nu}$ is the rotation tensor and $R_{\mu\nu}$ is the Ricci tensor. Since the timelike congruence that is orthogonal to the hypersurface foliating it satisfies $\omega^{\mu\nu}=0$~\cite{ah5}, we have the focusing of geodesics
\begin{equation}
    \frac{d\vartheta}{d\tau} \leqslant 0 \qquad \text{if} \qquad R_{\mu\nu}u^\mu u^\nu \geqslant 0,
\end{equation}
as $\sigma^{\mu \nu} \sigma_{\mu \nu} \geqslant 0$ for a timelike vector.

The weakest of the energy condition, the \textit{null energy condition} states that $T_{\mu\nu} k^\mu k^\nu \\geqslant 0$ for some null vector $k^\mu$. This inequality may be satisfied if one or all other conditions are violated. However, the violation of this condition implies the violation of all.

We can express the components the energy-momentum tensor at a space-time point $P$ with respect to an orthonormal basis $e_{\hat a}$, $\hat a=\{0,1,2,3\}$ by using the relation $T_{\hat{a}\hat{b}}=T_{\mu\nu}e_{\hat a}^\mu e_{\hat b}^\nu$. Doing so gives the four general form of the orthonormal energy-momentum tensor $T_{\hat{a}\hat{b}}$ forming the Hawking-Ellis~\cite{10,ec20} classification scheme (we present a more fine-grained Segre classification~\cite{es22} in the type \Romannum{3} Hawking-Ellis form which will be a basis for the classification of the Kerr-Vaidya metrics). Categorizing the energy-momentum tensor involves solving the Lorentz-invariant eigenvalue problem~\cite{ec20}
\begin{equation}
    |T^{\hat{a}\hat{b}}-\lambda \eta^{\mu\nu}|=0,
\end{equation}
where $\eta^{\mu\nu}$ is the Minkowskian metric and $\lambda$ is the eigenvalue. The four different forms of the energy-momentum tensor in the orthonormal basis are:

\textbf{Type \Romannum{1}:} \begin{equation}
        T^{\hat a \hat{b}}= \left(\begin{tabular}{c|ccc}
$\rho$ & 0  & 0 & 0\\ \hline
0 & $p_1$  & 0 & 0 \\
0 & 0 & $p_2$ & 0 \\
0 & 0 & 0 & $p_3$
 \end{tabular}\right).
    \end{equation}
This form of the energy-momentum tensor has one timelike and three spacelike eigenvectors (constituting at least one timelike invariant two-plane mapped to itself by $T^{\hat a}_{\hat{b}}$) with four eigenvalues $-\rho$, $p_1$, $p_2$ and $p_3$. Many classical massive and massless fields and semiclassical fields in equilibrium have this form.

The type \Romannum{1} tensor satisfies the null energy condition if $\rho+p_i\geqslant0$ where, $i=\{1,2,3\}$. Similarly, it satisfies the weak energy condition if $\rho+p_i\geqslant0$ and $\rho\geqslant0$. Moreover, it satisfies the strong energy condition if $\rho+p_i\geqslant0$ and $\rho+\sum p_i\geqslant0$.

\textbf{Type \Romannum{2}:} \begin{equation}
        T^{\hat a \hat{b}}= \left(\begin{tabular}{c|ccc}
$\rho+f$ & $f$  & 0 & 0\\ \hline
$f$ & $-\rho+f$  & 0 & 0 \\
0 & 0 & $p_2$ & 0 \\
0 & 0 & 0 & $p_3$
 \end{tabular}\right). \label{t21.48}
    \end{equation}
This form of the energy-momentum tensor is the special case with one double null eigenvector $k^{\hat a} =(1,1,0,0)$ (constituting at least one timelike invariant two-plane mapped to itself by $T^{\hat a}_{\hat{b}}$). They are the characteristics of radiation field/null fluid traveling in the direction of $k^{\hat a}$.

The type \Romannum{2} tensor satisfies the null energy condition if $f\geqslant0$ and $\rho+p_i\geqslant0$ where, $i=\{2,3\}$. Similarly, it satisfies the weak energy condition if $f\geqslant0$, $\rho+p_i\geqslant0$ and $\rho\geqslant0$. Moreover, it satisfies the strong energy condition if $f\geqslant0$, $\rho+p_i\geqslant0$ and $\sum p_i\geqslant0$.

The energy-momentum tensor of the Vaidya space-times belong to type \Romannum{2}. The only non-zero component of the energy-momentum tensor of the Vaidya metrics (in both advanced and retarded coordinates) is $T_{oo}=\pm\frac{M'(o)}{4\pi r^2}$, where the positive sign is for $o=v$, and the negative sign is for $o=u$. We use the following tetrads to represent the energy-momentum tensor in an orthonormal basis:
\begin{align}
    &e_{\hat 0}= \left(-1,\pm M/r,0,0\right), \qquad e_{\hat 1}= \left(1,\pm\left(1-M/r\right),0,0\right), \nonumber \\
    &e_{\hat 2}=\left(0,0,1/r,0\right), \qquad
    e_{\hat 3}=\frac{1}{r}\left(0,0,0,\csc\theta\right),
\end{align}
where $+/-$ sign is for advanced/retarded coordinates. These tetrads satisfiy the following orthonormality and completeness relations:
\begin{align}
    & e_{\hat a}^\mu e_{\hat b \mu}=0 \qquad \textrm{for} \qquad \hat a \ne \hat b \nonumber\\
    & e_{\hat 0}^\mu e_{\hat 0 \mu}=-1, \nonumber\\
    &e_{\hat a}^\mu e_{\hat a \mu}=1 \qquad \textrm{for} \qquad \hat a \ne 0.
\end{align}
So, the components of the energy-momentum tensor in the orthonormal basis can be written as $T_{\hat{a}\hat{b}}=T_{\mu\nu}e_{\hat a}^\mu e_{\hat b}^\nu$, and this explicitly is:
\begin{equation}
    T^{\hat a \hat{b}}= \left(\begin{tabular}{cc|cc}
 $T_{0 0}$ & $T_{0 0}$  & 0& 0\\
$T_{0 0}$ & $T_{0 0}$  & 0& 0 \\ \hline 
0 & 0 & 0 & 0 \\
0 & 0 & 0 & 0
 \end{tabular}\right).
 \end{equation}
This orthonormal energy-momentum tensor is of the type \Romannum{2} form given in Eq.~\ref{t21.48} with $\rho=p_2=p_3=0$ and $f=T_{oo}$. This form of the energy-momentum tensor satisfies the null energy condition if $f\geqslant0$ and thus $M'(v)\geqslant0$ for the Vaidya metric in advanced coordinates and $M'(u)\leqslant0$ for the Vaidya metric in retarded coordinates.

\textbf{Type \Romannum{3}:} \begin{equation}
        T^{\hat a \hat{b}}= \left(\begin{tabular}{c|ccc}
$\rho$ & 0  & $f$ & 0\\ \hline
0 & $-\rho$  & $f$ & 0 \\
$f$ & $f$ & $-\rho$ & 0 \\
0 & 0 & 0 & $p$
 \end{tabular}\right).
    \end{equation}
This form of the energy-momentum tensor has a triple null eigenvector $k^{\hat a} =(1,1,0,0)$ (constituting at least one null invariant two-plane mapped to itself by $T^{\hat a}_{\hat{b}}$) with the eigenvalues $-\rho$, $-\rho$, $-\rho$ and $p$. The fine-grained Segre classification scheme~\cite{es22} can be used to divide type \Romannum{3} form into two more subclasses: 1) The eigenvalues $\rho$ and $p$ are distinct, that is, $\rho\ne p$ and 2) All four eigenvalues are equal, that is, $\rho=p$. Both these subforms could violate all the energy conditions, including the null energy condition, unless $f=0$, in which case it reduces to type \Romannum{1}.

\textbf{Type \Romannum{4}:} \begin{equation}
        T^{\hat a \hat{b}}= \left(\begin{tabular}{c|ccc}
$\rho$ & $f$  & 0 & 0\\ \hline
$f$ & $-\rho$  & 0 & 0 \\
0 & 0 & $p_1$ & 0 \\
0 & 0 & 0 & $p_2$
 \end{tabular}\right).
    \end{equation}
This form of the energy-momentum tensor has no real timelike or null eigenvectors but constitutes at least one timelike invariant two-plane mapped to itself by $T^{\hat a}_{\hat{b}}$. Like type \Romannum{3}, it also violates all the energy conditions including the null energy condition unless $f=0$, in which case it reduces to the type \Romannum{1}.

\section{The Petrov Classification}

The set of four null tetrads, for example, Eq.~\eqref{nt121}, satisfying the orthogonality and completeness relations given in Eq.~\eqref{nj125} can be used as a basis to represent the Einstein equation. This alternate procedure of tackling the problems of general relativity forms the Newman-Penrose formalism~\cite{np73}. When the components of tensors are expressed on these bases, the second order differential equations of general relativity get converted into the first order and the number of the equations get reduced to half (because of the use of a complex basis).  These tetrads induce the constant symmetric metric Eq.~\eqref{np119}
\begin{equation}\label{20}
    \eta_{\mu \nu}
    =
    \begin{pmatrix}
    0 & 1 & 0 & 0\\
    1 & 0 & 0 & 0\\
    0 & 0 & 0 & -1\\
    0 & 0 & -1 & 0
   \end{pmatrix}.
\end{equation}
This metric allows to introduce the Weyl tensor as a part of the decomposition of the Riemann tensor into the irreducible representation of the Lorentz group
\begin{equation}
    R_{\alpha\beta\gamma\delta}=C_{\alpha\beta\gamma\delta}+ E_{\alpha\beta\gamma\delta}+ G_{\alpha\beta\gamma\delta},\label{np150}
\end{equation}
where
\begin{align}
    &E_{\alpha\beta\gamma\delta}=\frac{1}{2}\left(\eta_{\alpha \gamma}R_{\beta \delta}-\eta_{\beta \gamma}R_{\alpha \delta}-\eta_{\alpha \delta}R_{\beta \gamma}+\eta_{\beta \delta}R_{\alpha \gamma}\right)
    +\frac{1}{12}\left(\eta_{\alpha \gamma}\eta_{\beta \delta}-\eta_{\alpha \delta}\eta_{\beta \gamma}\right)R, \nonumber\\        &G_{\alpha\beta\gamma\delta}=\frac{1}{12}\left(\eta_{\alpha \gamma}\eta_{\beta \delta}-\eta_{\alpha \delta}\eta_{\beta \gamma}\right)R.
\end{align}
The Weyl conformal tensor $C_{\alpha\beta\gamma\delta}$ is the trace-free part of the curvature tensor and has ten independent components. The Newman-Penrose formalism allows us to represent these ten components by five scalars defined in the following manner
\begin{align}
    &\Psi_0=C_{\alpha \beta \gamma \delta}l^\alpha m^\beta l^\gamma m^\delta,\qquad
    \Psi_1=C_{\alpha \beta \gamma \delta}l^\alpha n^\beta l^\gamma m^\delta, \qquad
    \Psi_2=C_{\alpha \beta \gamma \delta}l^\alpha m^\beta \tilde m^\gamma n^\delta,\nonumber\\
    &\Psi_3=C_{\alpha \beta \gamma \delta}l^\alpha n^\beta \tilde m^\gamma n^\delta, \qquad
    \Psi_4=C_{\alpha \beta \gamma \delta}n^\alpha \tilde m^\beta n^\gamma\tilde m^\delta.
    \label{np152}
\end{align}
Being the trace-free tensor $C_{\alpha\beta\gamma\delta}$ allows one to define
\begin{equation}
    Q_{\alpha\gamma}=-C_{\alpha\beta\gamma\delta}^* u^\beta u^\delta,
    \label{np153}
\end{equation}
for any timelike vector $u^\alpha$ satisfying $u^\alpha u_\alpha=-1$, where
\begin{equation}
    C_{\alpha\beta\gamma\delta}^*=C_{\alpha\beta\gamma\delta}+\frac{i}{2}\epsilon_{\alpha\beta\mu\nu}C^{\mu\nu}_{~~~\gamma\delta}=C_{\alpha\beta\gamma\delta}+\frac{i}{2}\epsilon_{\gamma\delta \mu\nu}C_{\alpha\beta}^{~~~\mu\nu},
\end{equation}
and $\epsilon_{\alpha\beta\mu\nu}$ is the Levi-Civita four-form that satisfies $\epsilon_{\alpha\beta\mu\nu}l^\alpha n^\beta m^\gamma \tilde m^\delta=-i$. $Q_{\alpha\beta}$ has the following properties:
\begin{equation}
    Q^\alpha_\alpha=0, \qquad Q_{\alpha\beta}=Q_{\beta\alpha}, \qquad Q_{\alpha\beta}u^\beta=0.
\end{equation}
The first two of these relations allows us to consider $Q_{\alpha\beta}$ as a complex symmetric $3\times3$-matrix $\bf{Q}$. Using Eq.~\eqref{np152} and Eq.~\eqref{np153}, the components of $\bf{Q}$ can be written explicitly as
\begin{equation}
    \bf{Q}
    =
    \begin{pmatrix}
    \Psi_2-\frac{1}{2}(\Psi_0+\Psi_4) & \frac{i}{2}(\Psi_4-\Psi_0) & \Psi_1-\Psi_3\\
    \frac{i}{2}(\Psi_4-\Psi_0) & \Psi_2+\frac{1}{2}(\Psi_0+\Psi_4) & i(\Psi_1+\Psi_3)\\
    \Psi_1-\Psi_3 & i(\Psi_1+\Psi_3) & -2\Psi_2
   \end{pmatrix}.
\end{equation}

The invariant characterization of the gravitational field is independent of the coordinate systems and involves the investigation of the algebraic structure of tensors $C_{\alpha\beta\gamma\delta}$ and $E_{\alpha\beta\gamma\delta}$ of Eq.~\eqref{np150}. The energy-momentum tensor is equivalent to $E_{\alpha\beta\gamma\delta}$ (via Einstein equations) and its classification is presented in Sec.~\ref{e11}. The classification of $C_{\alpha\beta\gamma\delta}$ is equivalent to solving the eigenvalue equation ${\bf{Q} \bf{r}}= \lambda \bf{r}$, where, $\bf{r}$ constitute the eigenvectors and $\lambda$ the eigenvalues of $\bf{Q}$. The characteristic equation corresponding to this eigenvalue problem is $|{\bf Q}-\lambda \bf{I}|=0$, where $\bf{I}$ is the identity matrix. This equation determines the order of the elementary divisor $(\lambda-\lambda_1)^{m_1}, ...\, , (\lambda-\lambda_n)^{m_n}$, $m_1+ ...\,+m_n=3$ belonging to the eigenvalues $\lambda_1, ...\, \lambda_n$.

The algebraic structure of $\bf{Q}$ based on the elementary divisor and the multiplicities of the eigenvalues of characteristic equations constitute the Petrov classification. The full list of the Petrov classification can be found in Refs.~\cite{ch41,es22}. We are here only concerned with the two classes: Petrov type \Romannum{2} (to which, we will show below that, the Kerr-Vaidya metrics belong) and Petrov type D (to which all the spherically symmetric and the Kerr space-times belong). The Petrov type \Romannum{2} field has the eigenvalues $\lambda_1=\lambda_2=\lambda/2$ corresponding to the eigenvector $\left(1/2,-i/2,0\right)$ and $\lambda_3=\lambda$ corresponding to $(0,0,1)$, giving rise to the following form of the Weyl's tensor
\begin{equation}
    \bf{Q}
    =
    \begin{pmatrix}
    1-\frac{\lambda}{2} & -i & 0\\
    -i & -\frac{\lambda}{2}-1 & 0\\
    0 & 0 & \lambda
   \end{pmatrix}.
\end{equation}
Moreover, a principal null vector $l^\mu$ exists for the space-time of Petrov \Romannum{2}, such that,
\begin{equation}
    C_{\alpha\beta\gamma\delta}l_\nu l^\beta l^\gamma-C_{\alpha\beta\gamma\nu}l_\delta l^\beta l^\gamma=0 \qquad \text{or} \qquad \Psi_0=\Psi_1=0.
\end{equation}
The space-time is called Petrov type D if there exist two principal null vectors $l^\mu$ and $n^\mu$ such that the scalars $\Psi_0=\Psi_1=0$ in the direction of $l^\mu$ and $\Psi_3=\Psi_4=0$ in the direction of $n^\mu$, simultaneously. The Petrov type D space-times allow an additional fourth constant of motion such that the problem of solving the geodesic equations can be reduced to one involving quadratures~\cite{ch41}. This might not necessarily be true for Petrov type \Romannum{2} space-times.

\let\cleardoublepage\clearpage

\chapter{Self-consistent approach}

Using the self-consistent approach, space-time near the trapped region has been studied in spherical symmetry in Refs.~\cite{17,18,st70}. We will first review this study outlining the relevant properties of the space-time close to the trapped region. The self-consistent approach assumes the validity of the semiclassical Einstein equation $G_{\mu\nu}=8\pi \left<T_{\mu\nu}\right>$ with the renormalized stress-energy tensor $\left<T_{\mu\nu}\right>$ representing the sum total of the classical matter field and quantum excitations. Besides, this scheme makes the following two assumptions: 1) Finite-time formation of the trapped region, and 2) The regularity of the boundary enclosing the trapped region, as expressed by finite values of the curvature scalars. However, this approach does not consider the particular excitation state of quantum fields and the existence of the Hawking radiation and the event horizon.

\section{Space-time near the trapped region in spherical symmetry}\label{s3}

The most general spherically symmetric space-time contains at most two independent parameters and these two parameters are indeed constrained by the two assumptions of the self-consistent approach~\cite{17,19}. Throughout this section, we mention only the horizon for an apparent horizon.

The general metric describing spherically symmetric space-time in Schwarzschild coordinates is
\begin{equation}
   ds^2=-e^{2h(t,r)}f(t,r)dt^2+{f(t,r)}^{-1}dr^2+r^2d\Omega^2,
   \label{eq1}
\end{equation}
where
\begin{equation}
   f(t,r)=1-\frac{C(t,r)}{r},
   \label{eq2}
\end{equation}
is an invariant quantity called Misner-Sharp invariant~\cite{20} and $e^{2h(t,r)}$ serves as an integration factor for coordinate transformations, for example
\begin{equation}
    dt=e^{-h}\left(e^{h_{\pm}}du_\pm \mp f^{-1} dr\right),
\end{equation}
where the upper sign is for advanced null and the lower sign is for retarded null coordinates. This transformation gives the metric of the form
\begin{equation}
    ds^2=-e^{2h_{\pm}(u_\pm,r)}f(u_\pm,r)dt^2\pm 2 e^{h_{\pm}(u_\pm,r)} du_\pm dr+r^2d\Omega^2.
    \label{nc24}
\end{equation}
For Schwarzschild space, $h=0$ and $C=2M$, $M$ being the mass of the black-hole. A trapped region exists in such space-time if $f(t,r_g)=0$ for some $r=r_g(t)$ and its largest root gives an apparent horizon~\cite{20}. Now, at the apparent horizon, the metric of Eq.~\eqref{eq1} is diverging. The results below are established in an attempt to construct finite invariant quantities from the divergent metric.

The two invariants constructed from the energy-momentum tensor, namely its trace $T^{\mu}_{\mu}=g^{\mu\nu}T_{\mu\nu}$ and its square ${\cal{K}}=T^{\mu\nu}T_{\mu\nu}$ are used to express the regularity of an apparent horizon. If these two invariants are finite, then the rest are finite at the apparent horizon~\cite{st70}. The only non-vanishing components of $T_{\mu\nu}$ in spherical symmetry are $T_{tt}$, $T^{rr}$, $T^r_{t}$, $T^{\theta}_{\theta}$, $T^{\phi}_{\phi}$. Therefore,
\begin{equation}
    T^{\mu}_{\mu}=-e^{-2h}T_{tt}/f+T^{rr}/f+2T^{\theta}_{\theta},
\end{equation}
and
\begin{equation}
    {\cal{K}} = -2(e^{-h}T^r_{t}/f)^2+(e^{-2h}T_{tt}/f)^2+(T^{rr}/f)^2+2(T^{\theta}_{\theta})^2,
\end{equation}
where $T^{\theta}_{\theta}=T^{\phi}_{\phi}$ for spherical symmetry. Further, the finiteness of $T^{\theta}_{\theta}$ in general relativity follows from the consistency of Einstein equations~\cite{17}. Now, the regularity condition implies that the functions $e^{-h}T^r_{t}$, $e^{-2h}T_{tt}$ and $T^{rr}$ scales as $f^k$ for some $k$ as $r\to r_g$. It is shown in Ref.~\cite{dt72} that only the solutions with $k=0,1$ satisfy the regularity conditions and the generic case corresponds to
\begin{equation}
    \lim_{r\to r_g}e^{-2h}T_{tt}=\lim_{r\to r_g}T_{rr}=\pm \lim_{r\to r_g}e^{-h}T^r_{t}=\pm\gamma(t)^2\leqslant 0.
\end{equation}
It is shown in Ref.~\cite{17} that the real solution only exists for $-\gamma^2$. This form of the energy momentum tensor, however, violates the null energy condition giving $T_{\mu\nu}k^{\mu}k^{\nu}<0$ for null vectors $k^{\mu}=(1, \pm1, 0, 0)$. This result is consistent with  the result of Ref.~\cite{10} that in an asymptotically flat space-time, the trapped surface is not accessible to a distant observer unless the weak energy condition is violated.

Now, for the $k=0$ solution, we have the explicit form of energy-momentum tensor near the horizon
\begin{equation}
    T_{tt}=-{\gamma}^2e^{2h}, \quad T^r_t=\pm{\gamma}^2e^h  \quad \textrm{and} \quad T^{rr}=-{\gamma}^2.
    \label{emt26}
\end{equation}
This generic form of the energy-momentum tensor is constrained solely on the assumption of spherical symmetry and regularity of the trapped region. Hence this result is expected to hold true in all metric theories of gravity~\cite{22}. This triple limit of the energy-momentum tensor was observer in the ab-initio calculation of the renormalized stress-energy tensor on the Schwarzschild background~\cite{27}. The energy-momentum tensor of the form given in Eq.~\eqref{emt26} gives the following metric functions on solving two of the three independent Einstein equations~\cite{17}
\begin{align}
    C(t,r)&=r_g(t)-4 \pi\gamma r_g^{3/2} \sqrt{r-r_g(t)}+{\mathcal{O}(x)}, \label{eqn19}\\
    h(t,r)&=-\frac{1}{2}\ln\frac{r-r_g(t)}{\xi_0(t)}+{\mathcal{O}(\sqrt{x})},\label{eqn20}
\end{align}
where $\xi_0(t)$ is a function that can be determined from the full solution of the Einstein equations. The third Einstein equation gives the consistency condition on the above solution
\begin{equation}
    \frac{r_g'}{\xi_0(t)}=\pm 4 \sqrt{\pi r_g}\gamma.
\end{equation}
It is evident from this equation that the positive (negative) sign represents the accreting (evaporating) case.

The metric Eq.~\eqref{eq1} with the coefficients given by Eq.~\eqref{eqn19} and Eq.~\eqref{eqn20} can be expressed conveniently in null coordinates (retarded for an accreting and advanced for an evaporating case). In the advanced null coordinates ($u_+ =v$), the metric coefficients of Eq.~\eqref{nc24} becomes~\cite{st70}
\begin{align}
    & C_+(v,r)=r_+(v)+ \sum_{i\geqslant1}w_i(v)(r-r_+)^i,\\
    & h_+(v,r)= \sum_{i\geqslant1}\chi_i(v)(r-r_+)^i,
\end{align}
where $w_i(v)\leqslant1$ and $\chi_i(v)$ are some functions. Only two out of the four possibilities of Vaidya metrics are consistent with the self-consistent solutions in spherical symmetry: $M'(v)<0$ in advanced coordinates and $M'(u)>0$ in retarded coordinates.

\subsection{Observers near the horizon}

The assumption of the formation of trapped region in finite time also allows us to extract some information on energy density and pressure even without knowing the equation of state. Consider a radially infalling observer with timelike 4-velocity $u^\mu=(\dot{T}, \dot{R}, 0, 0)$. The outgoing radial normal of $u^\mu$ is $n_\mu=e^h(-\dot{R}, \dot{T}, 0, 0)$, which is spacelike. The energy density and pressure in the frame of the infalling observer is given as $\rho=T_{\mu\nu}u^{\mu}u^{\nu}$ and $p=T_{\mu\nu}n^{\mu}n^{\nu}$. As the 4-velocity $u^{\mu}u_{\nu}=-1$ we have
\begin{equation}
    \dot T=\frac{\sqrt{F+\dot R^2}}{e^H F}, \label{eq26}
\end{equation}
where $H=h(T,R)$ and $F=f(T,R)$. Substituting the values of $T_{\mu\nu}$, $u^{\mu}$, $n^{\mu}$ onto $\rho$ and $p$, we thus obtain
\begin{equation}
    \rho=p=-\frac{\left(\dot{R}\pm\sqrt{F(T, R)+{\dot{R}}^2}\right)^2}{{F(T, R)}^2}{\gamma}^2,
\end{equation}
where the positive sign gives the density and pressure for the retreating horizon ($r_g'<0$) and the negative sign gives their values for the advancing horizon ($r_g'>0$).

In the limit $R\to r_g$, we have
\begin{align}
    \rho=p\approx-\frac{4{\dot{R}}^2{\gamma}^2}{F^2}  \quad \textrm{for} \quad r_g'>0, \label{eq20}\\
    \rho=p\approx-\frac{{\gamma}^2}{4\dot R^2}  \quad \textrm{for} \quad r_g'<0. \label{eq21}
\end{align}
where, Eq.~\eqref{eq2} and Eq.~\eqref{eqn19} gives $F^2 \sim 16 \pi^2\gamma^2 r_g (R-r_g)$. Similarly, the flux $\phi=T_{\mu\nu}u^{\mu}n^{\nu}$ is given by
\begin{equation}
    \phi=\pm\rho,
    \label{eq22}
\end{equation}
where the positive sign is for the retreating horizon and the negative sign is for the advancing horizon. From Eq.~\eqref{eq20} and Eq.~\eqref{eq22}, we can see that the expanding trapped region is characterized by the the divergence of energy density, pressure and flux as $f\to 0$ for $r\to r_g$. However, this divergence does not accompany a singularity $r_g(T)$ as the invariants $T^{\mu}_{\mu}$ and $\cal{K}$ are finite there. Thus, an infalling observer encounters a finite value of energy density, pressure, and flux near the horizon of an evaporating black hole $r_g'<0$. However, the observer encounters a firewall at the horizon of an accreting black hole $r_g'>0$.

\subsection{Violation of the quantum energy inequality} \label{113}

Quantum energy inequality gives the limit on the allowed limiting value of the local energy density averaged along the timelike curve or along the timelike sub-manifold. Let $\xi$ be the smooth timelike world-line and $\mathscr{F}$ be the real-valued function with the compact support. Then the quantum energy inequality for the space-time with small curvature along the geodesic $\xi$ can be written as~\cite{23}
\begin{equation}
    \int_\xi {d\tau \mathscr{F}^2(\tau)\rho(\tau)} \geqslant {\cal B}_A,
    \label{eq23}
\end{equation}
where ${\cal B}_A$ is a bound function depending on $\mathscr{F}$, $\xi$ and the Ricci scalar $8\pi T^\mu_\mu$ of the space-time.

We now consider an advancing horizon with $r_g'>0$ for which the energy density, pressure and the flux is given by Eq.~\eqref{eq20}. In the semiclassical limit, the curvature near the horizon is small enough such that Eq.~\eqref{eq23} is applicable and the horizon radius $r_g$ does not change appreciably such that $\dot{R}- \dot{r}_g\approx \dot{R}$. We now consider $\mathscr{F}(\tau)=1$ at the horizon crossing and $\mathscr{F}(\tau)=0$ in the domain of the violation of null energy condition, which gives
\begin{equation}
    \int_\xi {d\tau \mathscr{F}^2(\tau)\rho(\tau)} \approx \int_\xi {- d\tau \frac{4{\dot{R}}^2{\gamma}^2}{f^2}}\approx\int_\xi {d\tau \frac{4 {\gamma}^2 r_g^2 |\dot{R}| dR}{ \alpha^2 (R-r_g)}},
\end{equation}
close to the horizon $R\to r_g$ for the radially infalling observer. Now, taking $|\dot{R}|\sim \textrm{constant}$, which is valid for the short trajectory, we get
\begin{equation}
    \int_\xi {d\tau \mathscr{F}^2(\tau)\rho(\tau)}\propto \ln(R-r_g)\sim -\infty,
\end{equation}
thereby showing the violation of a quantum energy inequality implying that the horizon can not grow. The only other possibility is that the semiclassical analysis is not valid close to the horizon.

\subsection{Horizon crossing by test particles}\label{114}

The massive test particle will cross the horizon if the distance $R(\tau)-r_g(T(\tau))\to0$. Crossing of the horizon does not happen if $\dot{R}-r_g'\dot{T}>0$ for $R(\tau)-r_g(T(\tau))>0$ where, $r_g'<0$ for an evaporating black-hole. Here we assume a quasi-stationary evaporation, so that, the leading order expansions of Eq.~\eqref{eq26} for $\dot{T}$ could give the correct result qualitatively. This assumption also enables us to neglect the backreaction on the metric because of the evaporation. We thus have
\begin{equation}
    \dot{R}-r_g'\dot{T}\approx\dot{R}-r_g'\frac{\dot{R} e^{-H}} {F}\bigg(1+\frac{F}{2\dot{R}^2}\bigg).
\end{equation}
Now, substituting the values of $F$ and $H$ from Eq.~\eqref{eqn19} and Eq.~\eqref{eqn20}, we get
\begin{equation}
    \dot{R}-r_g'\dot{T} \approx-\frac{(\dot R^2-4\pi r_g^2 \gamma^2)}{2|\dot R|\sqrt{\pi}r_g^{3/2}\gamma}\sqrt{R-r_g},
    \label{hc218}
\end{equation}
where quasi-stationary evaporation law has been assumed in obtaining this. For such black holes, we can take the limit $r_g' \approx -\kappa/r_g^2$ where, the surface gravity $\kappa \sim 10^{-3}-10^{-4}$~\cite{26,27} and this gives~\cite{19}
\begin{equation}
    \xi_0=\frac{1}{2}\sqrt{\frac{2\kappa}{r_g}}.
    \label{hc219}
\end{equation}
Hence, from Eq.\eqref{hc218} and Eq.\eqref{hc219}, horizon crossing by the test particle is not possible if:
\begin{equation}
    \begin{split}
        |\dot{R}|<2\sqrt{\pi}r_g\gamma= \sqrt{\frac{\kappa}{r_g^2}}\approx \frac{0.01}{r_g},
    \end{split}
\end{equation}
and this gives the numerical estimate for which horizon crossing is prevented for a macroscopic non-rotating black-hole. This difficulty in crossing the horizon by slowly-moving test particle is consistent with the results in existing literature~\cite{28}.

The same lines of calculations follow for massless test particles approaching the horizon but with the trajectory parametrized by some arbitrary non-zero $\lambda$. The parametrization will be most convenient if we choose $\lambda=-R$ and, in that case, $d\lambda/dR=-1$, thereby showing that massless particles always cross the apparent horizon.

\section{Kerr-Vaidya line-element}

As astrophysical objects are naturally rotating, using the above results obtained for spherical symmetry might be an oversimplification of real physics. Similarly, the firewall of the horizon of an accreting black hole might be an artifact of spherical symmetry. Hence, this project is an attempt to generalize the results by extending the research in axisymmetric space-time. So, we aim on studying the properties of space-time near an apparent horizon in axial symmetry based on the above-mentioned, two assumptions of the self-consistent approach: finite time formation of an apparent horizon and the regularity of it.

Application of the self-consistent approach to the axial symmetry requires us to treat the general time-dependent axisymmetric metric that has seven parameters. These parameters depend on three variables as compared to the general spherical metric with two parameters depending on two variables~\cite{ch41}
\begin{equation}
    d{\tau}^2=-\frac{f}{\kappa^2}{dt}^2+\kappa^2 r^2 \sin^2{\theta}\left(d{\phi}-q_2{dr}-q_3{d\theta}-\frac{\omega}{r\kappa^2}{dt}\right)^2+\sigma\left(\frac{\beta^2}{f}dr^2+r^2 d{\theta}^2\right),
\end{equation}
where the parameters $f$, $\beta$, $\sigma$, $\kappa$, $\omega$, $q_2$ and $q_3$ depend on the coordinates $t$, $r$ and $\theta$. Solving the second-order non-linear Einstein equation is enormously complex in this setup.

We thus aim to simplify the problem by working on the Kerr-Vaidya metric (in both advanced and retarded coordinates) to see if the results of spherical symmetry can be generalized. As the Kerr-Vaidya metric reduces to Kerr metric (an asymptotic black-hole solution in axial symmetry) in the stationary case, our result is expected to be sufficiently general to study the black-holes occurring in nature which are modeled as Kerr black-holes. Moreover, Kerr-Vaidya metrics in advanced/retarded coordinates are also the generalization of the Vaidya metrics in advanced/retarded coordinates which are the leading order solution of the evaporating/accreting black holes in axial symmetry. Having a taste of what the results look like in axial symmetry, we then plan to extend our research to study the general axial symmetric metric with a reduced number of parameters.

To see that Kerr-vaidya metrics are indeed the axisymmetric analogue of the Vaidya metrics, we allow mass $M$ of the Kerr metric in retarded coordinates given in Eq.~\eqref{ook127} to be the function of retarded time $u$. We then obtain the Kerr-Vaidya metric in retarded coordinates~\cite{39}
\begin{dmath}
ds^2=-\bigg(1-\frac{2 M(u) r}{\rho^2}\bigg)du^2-2 du dr+\rho^2 d\theta^2-
    \frac{4 a M(u) r \sin^2\theta}{\rho^2}d\phi du
    +2 a \sin^2\theta d\phi dr+
    \frac{(r^2+a^2)^2-a^2 \Delta \sin^2\theta}{\rho^2}\sin^2\theta d\phi^2,
    \label{k47}
\end{dmath}
where $\Delta=r^2-2 M(u) r+a^2$ here. The Kerr-Vaidya metric in advanced coordinates can similarly be obtained.

\subsection{Kerr-Vaidya metric as a Newman-Janis transformation of the Vaidya metric}

The Newman-Janis transformation (Sec.~\ref{nj12}) can be extended to obtain the radiating black hole solution in axial symmetry from the radiating solution in spherical symmetry. There have been numerous works in the past to obtain the rotating regular black hole solution and rotating radiating regular black hole solution using this prescription, all in retarded coordinates~\cite{nj2,nj3,nj4}. Here we derive a radiating rotating black hole solution in advanced coordinates using the complex coordinate transformation suggested by Newman and Janis given the importance of these coordinates to study the evaporating black holes.

We start by considering the Vaidya metric in advanced coordinates as the seed metric
\begin{equation}
    ds^2=-f(v,r) dv^2+2 dv dr+r^2 d\theta^2 +r^2 \sin^2\theta d\phi^2,
\end{equation}
where $f(v,r)=1-2 M(v)/r$ is an arbitrary function. This metric can be expressed in terms of the complex null tetrad as in Eq.~\ref{np119}. However, two real and two complex null vectors, in this case, are
\begin{equation}
    \begin{split}
        & l^\mu=\delta^\mu_v+ \frac{1}{2}f(v,r) \delta^\mu_r, \qquad n^\mu=-\delta^\mu_r \qquad \text{and}\\
        & m^\mu=\frac{1}{\sqrt{2}r}\left(\delta^\mu_\theta+\frac{i}{\sin\theta}\delta^\mu_\phi \right), \qquad \tilde m^\mu=\frac{1}{\sqrt{2}r}\left(\delta^\mu_\theta-\frac{i}{\sin\theta}\delta^\mu_\phi \right).
    \end{split}
\end{equation}
These four vectors satisfy the same completeness and orthogonality relations given in Eq.~\eqref{nj125}. In analogy to the Newman-Janis prescription, we perform the following complex coordinate transformation:
\begin{equation}
    {x'}^\mu=x^\mu- i a \left(\delta^\mu_r+\delta^\mu_v \right),
\end{equation}
where the prime denotes the new coordinate, not the derivative. As a result, the tetrad $Z^\mu=\left(l^\mu, n^\mu, m_\mu, \tilde m_\mu \right)$ transforms as ${Z'}^\mu_i = \left(\frac{\partial {x'}^\mu}{\partial x^\nu}\right) Z^\nu_i$, thereby giving
\begin{equation}
    \begin{split}
        & {l'}^\mu=\delta^\mu_v+ \frac{1}{2}{\cal F} (v,r, \theta) \delta^\mu_r \qquad {m'}^\mu=\frac{1}{\sqrt{2}(r-i a \cos\theta)}\left(i a \left(\delta^\mu_v+ \delta^\mu_r \right)\sin\theta +\delta^\mu_\theta+ \frac{i}{\sin\theta} \delta^\mu_\phi \right)\\ & {n'}^\mu=-\delta^\mu_r \qquad \tilde {m'}^\mu=\frac{1}{\sqrt{2}(r+i a \cos\theta)}\left(-i a \left(\delta^\mu_v+ \delta^\mu_r \right)\sin\theta +\delta^\mu_\theta -\frac{i}{\sin\theta}\delta^\mu_\phi \right),
        \label{nt231}
    \end{split}
\end{equation}
where the new function ${\cal F} (v,r, \theta)$ depends on $f(v,r)$ through the complex coordinate transformation
\begin{equation}
    {\cal F} (v,r, \theta)= 1-\frac{2 M(v)}{2}\left(\frac{1}{r'}+\frac{1}{\tilde r'}\right)=1-\frac{2 M(v) r}{\rho^2}.
    \label{nj233}
\end{equation} 
Now, this transformed tetrad yields a new metric given by
\begin{equation}
    {g'}^{\mu\nu}=- {l'}^\mu {n'}^\nu- {l'}^\nu {n'}^\mu+ {m'}^\mu \tilde {m'}^\nu+ {m'}^\nu \tilde {m'}^\mu.
\end{equation}
We thus have the new line-element
\begin{dmath}
ds^2=-{\cal F} (v,r, \theta) dv^2+2 dv dr+\rho^2 d\theta^2+
    2 a \left(-1+{\cal F} (v,r, \theta)\right) \sin^2\theta d\phi dv
    -2 a \sin^2\theta d\phi dr
    + \left(r^2+a^2+ a^2 \left(1-{\cal F} (v,r, \theta)\right) \sin^2\theta \right)\sin^2\theta d\phi^2,
\end{dmath}
where $\rho^2=r^2+a^2 \cos^2\theta$. Substituting ${\cal F} (v,r, \theta)$ from Eq.~\eqref{nj233} in above expression gives the line element
\begin{dmath}
ds^2=-\bigg(1-\frac{2 M(v) r}{\rho^2}\bigg)dv^2+2 dv dr+\rho^2 d\theta^2-
    \frac{4 a M(v) r \sin^2\theta}{\rho^2}d\phi dv
    -2 a \sin^2\theta d\phi dr
    + \frac{(r^2+a^2)^2-a^2 \Delta \sin^2\theta}{\rho^2}\sin^2\theta d\phi^2.
    \label{k44}
\end{dmath}
This is the Kerr-Vaidya metric in advanced coordinates. Substantial works have been done in the past to show that this metric does indeed represents the radiating rotating solution of the Einstein equation~\cite{36,37,38}.


\let\cleardoublepage\clearpage

\chapter{Locating an apparent horizon of the Kerr-Vaidya metric}\label{ah31}

We have shown above that the apparent horizon of the Kerr black hole coincides with its event horizon. It is located at the largest root of $\Delta=0$, that is $r_0=\vcentcolon M+\sqrt{M^2-a^2}$.
We have also shown that for both the ingoing and the outgoing  Vaydia metrics the apparent horizon is located at $r_g\equiv r_0=2M$. For the  Kerr-Vaidya  metric of Eq.~\eqref{k47} the same relation $r_g(u)=r_0=M(u)+\sqrt{M^2(u)-a^2}$ holds~\cite{ah12}. It was shown in Ref.~\cite{39} that this is not true for the metric of Eq.~\eqref{k44} and in this case the difference $r_g(\theta)-r_0$ is of the order $|M'|$.

\section{Kerr-Vaidya metric in advanced coordinates}

The standard approach~\cite{ah11,ah70} for constructing the ordinary differential equation for the apparent horizon on $v=\text{constant}$ surface is based on exploiting properties of a spacelike foliation.  It cannot be used in this case as the foliating hypersurfaces are timelike. However, since the approximate location of the apparent horizon is known, we obtain the leading correction in $M'$ by using the methods of analysis of null congruences and hypersurfaces~\cite{ah5}.

\subsection{Outgoing null congruence orthogonal to the $v=\text{constant}$ surface}

Assume that at some advanced time $v$ the apparent horizon is located  at $r_g=r_0(v)+z(\theta)$, where the function $z(\theta)$ is to be determined. This two-surface has two linearly independent tangent vectors:
\begin{equation}
    t_1^\mu=e_3^\mu, \qquad t_2^\mu=\ z'(\theta) e_1^\mu+e_2^\mu, \label{h33}
\end{equation}
where $e^\mu_a =\frac{\partial x^\mu}{\partial y^a}$. Here, $x^\mu=(v, r, \theta, \phi)$ is the space-time coordinates and $y^a=(r, \theta, \phi)$ is the coordinates of the hypersurface. We obtain the outward- and inward-pointing future-directed null vectors $l^+\equiv l$ and $l^-\equiv n$ by using the orthogonality condition $l^\pm_\mu t^\mu_{1,2}=0$. Before the rescaling $l^v=1$ and the  normalization $n_\mu l^\mu=-1$ the two null vectors are given by
\begin{equation}
    l^\pm_\mu=(-1,\zeta_\pm , - \zeta_\pm z'(\theta),0). \label{h34}
\end{equation}
The two values of $\zeta_\pm$ are obtained from the null condition $l^\mu l_\mu=0$,
\begin{equation}
    \zeta_\pm=\frac{r^2+a^2\pm \sqrt{2 a^2 r M \sin ^2 \theta+\rho^2 \left(a^2+r^2\right)-a^2 (z'(\theta))^2\sin^2\theta}} {\Delta+ (z'(\theta))^2}.
\end{equation}
Vectors $l^\mu$ and $n^\mu$ can be extended to a field of tangent vectors of the affinely-parameterized null geodesics in the bulk. Once the future-directed outgoing null geodesic congruence orthogonal the surface is identified, calculating the expansion $\vartheta_l$ and equating it to zero results in the differential equation for $z(\theta)$.  There are at least two equivalent ways to calculate the expansion.

\subsection{Area expansion on the $v=\text{constant}$ surface}

One approach to calculation of the expansion uses its geometric meaning as a relative rate of change of the two-dimensional cross-sectional area  with an infinitesimal evolution of the geodesic from $v$ to $v+\delta v$, Eq.~\eqref{exp114}. For this, we consider a geodesic $\xi$ originating somewhere from the hypersurface $v=\text{constant}$ and we label it with an arbitrary parameter $\lambda=\lambda_P$. The hypersurface $v=\text{constant}$ can be defined by the coordinate system $(r, \theta, \phi)$ which induces the metric
\begin{equation}
    h_{a b}= g_{\mu \nu} e^\mu_a e^\nu_b,
    \label{46}
\end{equation}
where $e^\mu_a$ is defined in Eq.~\eqref{h33}. We also consider a small neighbourhood around $P$ such that: a) Through each of these neighbourhood points $P'$, there passes another geodesic from the congruence $l$, and b) At each point $P'$, the parameter $\lambda_{P'}=\lambda_P$.

We now take a two-surface $r=r_0+z(\theta)$ on the three hypersurface $v=\text{constant}$. Let us define a coordinate system on this two-surface $S$:
\begin{equation}
    e_{1P}=(0, 0, 1), \qquad e_{2P}=(z'(\theta), 1, 0).
\end{equation}
This two-surface $S$ evolves, as the geodesic $\xi$ on $v=\text{constant}$ hypersurface with $\lambda=\lambda_P$ evolves to the $v+\delta v=\text{constant}$ hypersurface with $\lambda=\lambda_Q$. Our goal here is to calculate the change in area of this two-surface $S$ as it evolves from $\lambda_P$ to $\lambda_Q$. The calculation would be very convenient if we choose the coordinate system comoving with the null geodesic $\xi$ to describe the two-surface $S$. In doing so, the coordinate system for $S$ at $\lambda=\lambda_Q$ can be written as
\begin{equation}
    e_{1Q}=(0, 0, 1), \qquad e_{2Q}=\left(z'(\theta)+ \partial_\theta l^2, 1+\partial_\theta l^3, 0\right).
\end{equation}
Now, the change in area of the two-surface $S$ as the geodesic $\xi$ propagates from point $P$ to $Q$ can be written as
\begin{equation}
    \delta A =\sqrt{\sigma_Q}d^2 y_Q -\sqrt{\sigma_P}d^2 y_P,
\end{equation}
where $y=(e_1, e_2)$ is the comoving coordinate that is constant along the geodesic and $\sigma=\det[\sigma_{i j}]$, $\sigma_{i j}$ being the two dimensional metric describing the two-surface $S$. $\sigma_{i j}$ can be calculated using the relation $\sigma_{i j}=e^a_i h_{a b} e^b_j$, where, $a,b=\{1,2,3\}$. As $y$ is constant as the geodesic propagates, we have $d^2 y_Q=d^2 y_P=d^2 y$, and thus
\begin{equation}
    \delta A =(\sqrt{\sigma_Q} -\sqrt{\sigma_P}) d^2 y.
\end{equation}
By the definition of an apparent horizon given in previous section, we have
\begin{equation}
    \delta A =(\sqrt{\sigma_Q} -\sqrt{\sigma_P}) d^2 y=0 \qquad \textrm{or} \qquad \sqrt{\sigma_Q} -\sqrt{\sigma_P}=0.
    \label{kvh311}
\end{equation}
Finally, direct calculation (under the assumption that $z(\theta)$ and its derivatives are of the order of $M'(v)$) yields (Appendix \ref{a1})
\begin{dmath}
8 r_0^2 \left(a^2+r_0^2\right) \left(\left(a^2+r_0^2\right) \left(\partial^2_\theta z(\theta)+\cot \theta \partial_\theta z(\theta)\right)-a^2 r_0 \sin ^2 \theta M' \right)
    +z(\theta) (r_0^2-a^2) \left(a^4+7 a^2 r_0^2+\left(r0^2- a^2\right) \cos 2 \theta +8 r_0^4\right)=0.
    \label{kvh313}
\end{dmath}
An alternative derivation is based on the direct evaluation of  $\vartheta_l=h^{\alpha\beta}l_{\alpha;\beta}$ on $S$, Eq.~\eqref{exp115}, which gives the same equation Eq.~\eqref{kvh313} (see Appendix \ref{a1} for the details).

The solution of the differential equation Eq.~\eqref{kvh313} gives the apparent horizon for the Kerr-Vaidya metric in advanced coordinates. Here, the term $\partial^2_\theta z(\theta)+\cot \theta \partial_\theta z(\theta)$ is the standard term occurring in the differential equation of the trapped surface of axisymmetric space-time in numerical relativity~\cite{ah11}. We solve this equation numerically to see the qualitative feature of the apparent horizon of the Kerr-Vaidya space-time in advanced coordinates.

\begin{figure}[!htbp]
\centering
\includegraphics[width=0.55\textwidth]{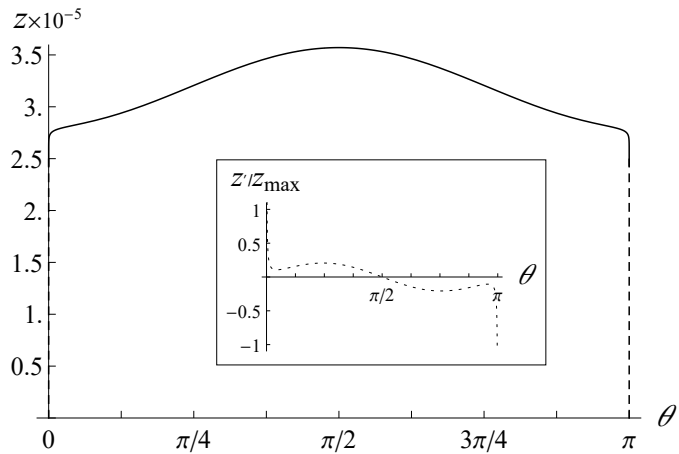}
\caption{Location of the apparent horizon relative to $r_0$ for $M=1$, $a=0.1$, $M'=-\kappa/M^2=0.01$. The equation was first solved as a boundary value problem $z(0)=0$, $z'(\pi/2)=0$,
resulting in $z_{\mathrm{m}}\vcentcolon= z(\pi/2)\approx3.57\times 10^{-5}$. The solution of the initial value problem $z(\pi/2)=z_{\mathrm{m}}$, $z'(\pi/2)=0$ coincides with the previous one
within the relative precision of $2\times 10^{-15}$ outside $\delta=10^{-6}$ interval from the poles.}
\label{f1}
\end{figure}

The graphical representation of the relative position of the apparent horizon from the horizon of the stationary Kerr metric obtained as a numerical boundary value problem is shown in Fig.~\ref{f1}.

From the Fig.~\ref{f1}, we can see that the trapped region coincides with the Kerr horizon $r_0$ at the pole and extends outside of it elsewhere (the maximum being at the equator). The assumption of $z'\ll 1$ fails near the poles, where $z=0$. The large value of $z'$ near the poles require additional checks to ensure that this is not the numerical artifact that goes beyond the validity of the initial approximation. This additional verification is done using the series solutions of Eq.~\eqref{kvh313} with the conventional initial conditions $z(0)=0$, $z'(0)$=0~\cite{ah11, ah70}, that is in the regime where the assumption $|z'|\ll 1$ is clearly valid,
\begin{equation}
    z_{\mathrm{ser}}= \frac{a^2 M' r_0}{16(a^2+r_0^2)}\theta^4+{\mathcal O}(\theta^5).
\end{equation}
For $M'<0$ it implies that at least near the poles $r_g>r_0$, that is at $r=r_0$ the expansion is still negative. However, this is impossible: at the poles the null congruence that is orthogonal to the two-dimensional surface $r=r_0$~\cite{39} has $\vartheta_l>0$. Moreover, using this solution to provide the initial values $z(\theta)$, $z'(\theta)$ at some $\theta=\epsilon\ll 1$  leads to inconsistencies.

We investigated the stability of this result in numerical experiments. For a fixed $M'=-\kappa/M^2$ the initial value problem $z(\pi/2)=z_0$,  $z'(\pi/2)=0$, where $z_0$ is some number,
leads to a well-behaved numerical solution.
However, the conditions $z(0)=z(\pi)=0$    are satisfied within a prescribed tolerance only for a very narrow range of the values $z_0$ around $z_{\mathrm{m}}=z(\pi/2)$ of the numerical solution of the above
 boundary value problem.

\section{For Kerr-Vaidya metric in retarded coordinates}

We use our method to give the simple derivation of the result that the apparent horizon of the retarded Kerr-Vaidya metric coincides with that of the stationary Kerr metric. Finding the null vector orthogonal to the $u=\text{constant}$ surface proceeds in exactly the same manner described above for the advanced coordinates. The future-directed null vector orthogonal to the $u=\text{constant}$ surface should be of the form
\begin{equation}
    l^\pm_\mu=(-1,\zeta_\pm , - \zeta_\pm z'(\theta),0),
\end{equation}
which on imposing the null condition $l^\mu l_\mu=0$ gives
\begin{equation}
    \zeta_\pm=-\frac{r^2+a^2\pm \sqrt{2 a^2 r M \sin ^2 \theta+\rho^2 \left(a^2+r^2\right)-a^2 (z'(\theta))^2\sin^2\theta}} {\Delta+ (z'(\theta))^2}.
\end{equation}
Unlike the advanced Kerr-Vaidya metric, the choice of the negative sign here gives the future directed outgoing null vector orthogonal to $u=\text{constant}$ surface. After identifying the outgoing null vector orthogonal to the surface, we need to calculate an area expansion as the vector evolves infinitesimally from $u=\text{constant}$ surface to $u+\delta u$. Again, the procedure is exactly the same as described above for the advanced coordinates. Direct calculation (under the assumption that $z(\theta)$ and its derivatives are of the order of $M'(u)$) yields $z(\theta)=0$ as an apparent horizon for the Kerr-Vaidya metric in retarded coordinates. Various authors, in the past, have shown that $r_0$ is the apparent horizon of the Kerr-Vaidya metric in retarded coordinates~\cite{nj3, ah12}.

\let\cleardoublepage\clearpage

\chapter{Space-time near the horizon of the Kerr-Vaidya metric}

\section{Null energy condition}

The energy-momentum tensor of both types of Kerr-Vaidya metrics can be represented schematically as (Appendix \ref{a2})
\begin{equation}
    T_{\mu\nu}=\begin{pmatrix}
    T_{oo}& 0 & T_{o\theta} &T_{o\phi} \\
    0 &0 &0 & 0 \\
    T_{o\theta} & 0 &0 & T_{\theta\phi} \\
    T_{o\phi} & 0 &T_{\theta\phi} & T_{\phi\phi} \end{pmatrix},\label{emt}
\end{equation}
where $o=u,v$. Using the null vector $w^\mu=(0,1,0,0)$, the energy-momentum tensor for both the metrics can be written concisely as
\begin{equation}
    T_{\mu \nu}=T_{0 0} w_\mu w_\nu + Q_\mu w_\nu + Q_\nu w_\mu,
\end{equation}
where
\begin{equation}
    Q_\nu=\bigg(0,\; 0,\; T_{0 \theta},\; \mp a\sin^2\theta\frac{r^2-a^2\cos^2\theta}{\rho^4}M' \bigg)
\end{equation}
where, $-/+$ sign corresponds to the advanced/retarded Kerr-Vaidya metrics. Further analysis of the energy conditions are simplified by using an orthonormal null tetrad to represent the energy-momentum tensor. We use a tetrad in which the null eigenvector $w^\mu=w^{\hat a}e^\mu_{\hat a}$ has the components $w^{\hat a}=(1,1,0,0)$, the third vector $e_{\hat 2}\propto \partial_\theta$ and the remaining vector $e_{\hat{3}}$ is found by completing the basis (Appendix \ref{a2}). The components of the energy-momentum tensor in the orthonormal basis can be written as $T_{\hat{a}\hat{b}}=T_{\mu\nu}e_{\hat a}^\mu e_{\hat b}^\nu$. Its explicit form is
\begin{equation}
    T^{\hat a \hat{b}}= \left(\begin{tabular}{cc|cc}
 $T_{0 0}$ & $T_{0 0}$  & $q_1$& $q_2$\\
$T_{0 0}$ & $T_{0 0}$  & $q_1$& $q_2$ \\ \hline 
$q_1$ & $q_1$ & 0 & 0 \\
$q_2$ & $q_2$ & 0 & 0
 \end{tabular}\right),
\end{equation}
where 
\begin{equation}
    q_1=-\frac{a^2 r M'}{8\pi \rho^5} \sin2\theta, \qquad q_2=\mp \frac{a(r^2-a^2\cos^2\theta) M'} {8\pi \rho^5}\sin\theta,
\end{equation}
Here, $-/+$ sign is for advanced/retarded metric. Using the null vector
\begin{equation}
    k_{\hat a}=(-1,\cos\alpha,\sin\alpha\cos\beta,\sin\alpha\sin\beta),
\end{equation}
the null energy condition becomes
\begin{equation}
    T_{0 0} (1-\cos\alpha)+2\sin\alpha(q_1 \cos\beta \pm q_2\sin\beta)\geqslant 0,
\end{equation}
where $+/-$ sign corresponds to the advanced/retarded metric. This identity always holds only when $T_{0 0} \geqslant 0$ and $q_1=q_2=0$. But, $q_1=q_2=0$ implies $a=0$ as $M' \ne 0$, in which case the metric reduces to the familiar Vaidya metric~\cite{35}. Hence, the null energy condition is always violated by the Kerr-Vaidya metric (in both advanced and retarded coordinates) when $a \ne 0$. Now, solving the eigenvalue equation
\begin{equation}
    |T^{\hat{a}\hat{b}}-\lambda \eta^{\mu\nu}|=0, \qquad \eta^{\mu\nu}=\mathrm{diag}(-1,1,1,1),
\end{equation}
gives the single quadrupled-degenerated eigenvalue $\lambda=0$, with the eigenvectors:
\begin{equation}
    w^{\hat a}=(1,1,0,0), \qquad Q^{\hat a}=\left(0,0,q_2/q_1,-1\right).
\end{equation}
The combination of the null eigenvector $w^{\hat a}$ and the spacelike eigenvector $Q^{\hat a}$ gives the null invariant plane of eigenspace. So, according to the classification scheme (Sec.~\ref{e11}), the energy-momentum tensor of the Kerr-Vaidya metrics belong to the type \Romannum{3}.

For the Petrov classification of the Kerr-Vaidya metric, we take the complex null tetrad given in Eq.~\eqref{nt123}, where $M=M(u)$. Here, we only present the calculation for the retarded Kerr-Vaidya metric. The calculation for the advanced Kerr-Vaidya metric proceed in the exactly same way (excpet that we should use the null tetrad of Eq.~\eqref{nt231} as a basis) and is of the same Petrov class. Now, we use Eq.~\eqref{np152} for the calculation of the Weyl scalars
\begin{align}
    &\Psi_0=0,\qquad
    \Psi_1=0, \qquad
    \Psi_2=\frac{M}{(r-i a \cos \theta)^3},\nonumber\\
    &\Psi_3=-\frac{a \sin\theta M' (a \cos\theta-3 i r)}{2 \sqrt{2} (r-i a \cos\theta)^3 (r+i a \cos\theta)}, \qquad
    \Psi_4=\frac{a^2 r \sin ^2\theta \left(M'' (a \cos\theta+i r)-2 i M'\right)}{2 (r-i a \cos\theta)^4 (a \cos\theta-i r)}.
\end{align}
We thus can see that retarded Kerr-Vaidya metric is of Petrov type \Romannum{2} and the vector ${l'}^\mu$ of Eq.~\eqref{nt123} is indeed the principal null geodesics of this space-time (also established in Ref.~\cite{pc74}).

\section{Local energy density}

\subsection{Kerr-Vaidya metric in advanced coordinates}

To calculate the energy density $\rho_0=T_{\mu\nu}u^\mu u^\nu$ perceived by an infalling observer, we assume the following four-velocity of an infalling observer
\begin{equation}
    u^\mu=(\dot{V},\; \dot{R},\; \dot{\Theta},\; \dot{\Phi}).
\end{equation}
We now choose the reference frame of the zero angular momentum observer which satisfies the condition $\partial_\phi^\mu u_\mu=0$, where, $\partial_\phi$ is the Killing vector. This gives:
\begin{equation}
    \dot{\Phi}=-\frac{g_{v\phi}\dot V+g_{r\phi}\dot R}{g_{\phi\phi}}=\frac{2 a M r\dot{V}+ a \rho^2 \dot{R}}{(r^2+a^2)^2- a^2 \Delta \sin^2\theta}.
\end{equation}
Thus, the four-velocity of the zero angular momentum observer up to the first order approximation in Taylor's expansion near $\Delta=0$ can be written as
\begin{equation}
    u^\mu=\left(\dot{V},\; \dot{R},\; \dot{\Theta},\;  \frac{2 a M r\dot{V}+ a \rho^2 \dot{R}}{(r^2+a^2)^2}\left(1+ \frac{a^2 \Delta \sin^2\theta}{(r^2+a^2)^2}\right)\right).
\end{equation}
Its covariant counterpart $u_\mu = g_{\mu\nu}u^\nu$ is
\begin{equation}
    u_\mu=\left(-\frac{\rho^2 \Delta}{ (r^2+a^2)^2} \dot{V}+\frac{\rho^2 \dot{R}}{r^2+a^2},\; \frac{\rho^2}{r^2+a^2}\left(\dot{V}-\frac{a^2 \sin^2\theta}{r^2+a^2}\dot{R}\right),\; \rho^2 \dot{\Theta},\; 0\right),
\end{equation}
where only the leading order coefficients of $\dot{V}$ and $\dot{R}$ are retained. The four-velocity being timelike satisfies $u^\mu u_\mu =-1$, from which one obtains
\begin{equation}
    -\frac{\rho^2 \Delta}{(r^2+a^2)^2}\dot{V}^2+ \frac{2\rho^2}{r^2+a^2}\dot{V}\dot{R}- \frac{\rho^2 a^2 \sin^2\theta}{(r^2+a^2)^2}\dot{R}^2+ \rho^2 \dot{\Theta^2}+1=0.
\end{equation}
This is the quadratic equation in $\dot V$, with two distinct roots. For an infalling observer $\dot R<0$, the condition $\dot V>0$ results in the unique solution
\begin{equation}
    \dot{V}\simeq -(\rho^2 \dot{\Theta^2}+1)\frac{r^2+a^2}{2 \rho^2 \dot{R}}+ \frac{a^2 \sin^2\theta}{2 (r^2+a^2)}\dot{R}.
    \label{v427}
\end{equation}
Now the local energy density for an infalling observer in the advanced Kerr-Vaidya metric can be calculated
\begin{dmath}
\rho_o =T_{v v}\dot{V^2}+2 T_{v \theta}\dot{V}\dot{\Theta}+ 2 (T_{v \phi}\dot{V}+ T_{\theta \phi}\dot{\Theta})\bigg(\frac{a \dot{V}}{r^2+a^2}-\frac{a \rho^2 \dot{R}}{(r^2+a^2)^2}\bigg)+ T_{\phi\phi}\bigg(\frac{a \dot{V}}{r^2+a^2}-\frac{a \rho^2 \dot{R}}{(r^2+a^2)^2}\bigg)^2\\
=\bigg(T_{v v}+ 2T_{v\phi}\frac{a}{r^2+a^2}+ T_{\phi\phi}\frac{a^2}{(r^2+a^2)^2}\bigg)\dot{V}^2+2 \bigg(T_{v \theta}+T_{\theta\phi}\frac{a}{r^2+a^2}\bigg)\dot{\Theta}\dot{V}+
\frac{2 a \rho^2}{(r^2+a^2)^2}\bigg(T_{v \phi}+\frac{a}{r^2+a^2}T_{\phi\phi}\bigg)\dot{V}\dot{R}+2T_{\theta\phi}\frac{a \rho^2}{(r^2+a^2)^2}\dot{\Theta}\dot{R}+ T_{\phi \phi}\frac{a^2 \rho^4}{(r^2+a^2)^4}\dot{R}^2,
\end{dmath}
where the explicit forms of the components of energy-momentum tensor $T_{\mu \nu}$ are given in Eqs.~\eqref{v41} - \eqref{v45} and $\dot{V}$ is given in Eq.~\eqref{v427}. Direct calculation  after substituting their values yields
\begin{align}
    \rho_0=&\frac{M'}{256 \pi \dot R^2 \left(a^2+r^2\right)^4 \left(a^2 \cos 2\theta+a^2+2 r^2\right)^2} \bigg(4 a^6 \dot \Theta^2+a^4 \left(\dot R^2+16 \dot \Theta^2 r^2+8\right)-\nonumber\\
    & a^4 \dot R^2 \cos 4\theta
+4 a^2 r^2 \left(\dot R^2+5 \dot \Theta^2 r^2+4\right)+4 a^2 \cos 2\theta \left(a^2 \dot \Theta+r (\dot R+\dot \Theta r)\right) \bigg(\dot \Theta \left(a^2+r^2\right)-\nonumber\\
&\dot R r\bigg) 
+8 \left(\dot \Theta^2 r^6+r^4\right)\bigg) \bigg(-2 a^6 \dot R^2+4 a^2 r^4 \left(4 a^2 \dot \Theta^2+3 \dot R^2+4\right)+4 r^6 \left(5 a^2 \dot \Theta^2+2\right) \nonumber\\
&+a^4 r^2 \left(4 a^2 \dot \Theta^2+7 \dot R^2+8\right)+a^2 \bigg(\dot R \left(a^2 \dot R \left(2 a^2+r^2\right) \cos 4\theta+16 \dot \Theta r \left(a^2+r^2\right)^2 \sin 2\theta\right) \nonumber\\
&+4 r^2 \cos 2\theta \left(a^4 \dot \Theta^2-2 a^2 (\dot R-\dot \Theta r) (\dot R+\dot \Theta r)-3 \dot R^2 r^2+\dot \Theta^2 r^4\right)\bigg)+8 \dot \Theta^2 r^8\bigg),
\end{align}
where the terms containing $M''$ has been neglected during the calculation as $M''\ll M'/M$ holds in the semiclassical limit. Near the equatorial plane, $\dot{\Theta}=0$, and this expression for the local energy density simplifies to
\begin{dmath}
\rho_0= \frac{\left(a^4+a^2 \left(\dot R^2+2\right) r^2+r^4\right) \left(a^4 \left(2 \dot R^2+1\right)+a^2 \left(3 \dot R^2+2\right) r^2+r^4\right) M'}{16 \pi \dot R^2 r^2 \left(a^2+r^2\right)^4}.
\end{dmath}
For $a=0$, we get the result for spherical symmetry derived in Ref.~\cite{19}.

\subsection{Kerr-Vaidya metric in retarded coordinates}

The calculation of the local energy density perceived by an infalling observer in this case proceeds in a similar manner as described above. For an infalling observer with four-velocity
\begin{equation}
    u^\mu=(\dot{U},\; \dot{R},\; \dot{\Theta},\; \dot{\Phi}),
\end{equation}
zero angular momentum condition $\partial_\phi^\mu u_\mu=0$, where, $\partial_\phi$ is the Killing vector, gives
\begin{equation}
    \dot{\Phi}=-\frac{g_{u\phi}\dot U+g_{r\phi}\dot R}{g_{\phi\phi}}=\frac{2 a M r\dot{U}- a \rho^2 \dot{R}}{(r^2+a^2)^2- a^2 \Delta \sin^2\theta}.
\end{equation}
Thus the four-velocity of the zero angular momentum observer up to the first order approximation in Taylor's expansion near $\Delta=0$ can be written as
\begin{equation}
    u^\mu=\left(\dot{U},\; \dot{R},\; \dot{\Theta},\;  \frac{2 a M r\dot{U}- a \rho^2 \dot{R}}{(r^2+a^2)^2}\left(1+ \frac{a^2 \Delta \sin^2\theta}{(r^2+a^2)^2}\right)\right).
\end{equation}
Its covariant counterpart $u_\mu = g_{\mu\nu}u^\nu$ is
\begin{equation}
    u_\mu=\left(-\frac{\rho^2 \Delta }{ (r^2+a^2)^2} \dot{U}- \frac{\rho^2 \dot{R}}{r^2+a^2},\; -\frac{\rho^2}{r^2+a^2}\left(\dot{U}+\frac{a^2 \sin^2\theta}{r^2+a^2}\dot{R}\right),\; \rho^2 \dot{\Theta},\; 0\right),
\end{equation}
where only the leading order coefficients of $\dot{U}$ and $\dot{R}$ are retained. The four-velocity being timelike should satisfy $u^\mu u_\mu =-1$, from which one obtains
\begin{equation}
    -\frac{\rho^2 \Delta }{ (r^2+a^2)^2} \dot{U}^2-\frac{2\rho^2}{r^2+a^2}\dot{U}\dot{R}- \frac{\rho^2 a^2 \sin^2\theta}{(r^2+a^2)^2}\dot{R}^2+ \rho^2 \dot{\Theta^2}+1=0.
\end{equation}
This is the quadratic equation in $\dot U$, with two distinct roots. For an infalling observer $\dot R<0$, the condition $\dot U>0$ results in the unique solution
\begin{equation}
    \dot{U} \simeq -\frac{2(r^2+a^2)}{\Delta}\dot{R}-\frac{r^2+a^2}{2 \rho^2 \dot{R}}\left(-\frac{\rho^2 a^2 \sin^2\theta}{(r^2+a^2)^2}\dot{R}^2+ \rho^2 \dot{\Theta}^2+1 \right). \label{u436}
\end{equation}
The local energy density for an infalling observer in Kerr-Vaidya metric in retarded coordinates can be calculated
\begin{dmath}
\rho_o = T_{u u}\dot{U^2}+2 T_{u \theta}\dot{U}\dot{\Theta}+ 2 (T_{u \phi}\dot{U}+ T_{\theta \phi}\dot{\Theta})\bigg(\frac{a \dot{U}}{r^2+a^2}-\frac{a \rho^2 \dot{R}}{(r^2+a^2)^2}\bigg)+ T_{\phi\phi}\bigg(\frac{a \dot{U}}{r^2+a^2}-\frac{a \rho^2 \dot{R}}{(r^2+a^2)^2}\bigg)^2\\
=\bigg(T_{u u}+ 2T_{u \phi}\frac{a}{r^2+a^2}+ T_{\phi\phi}\frac{a^2}{(r^2+a^2)^2}\bigg)\dot{U}^2+2 \bigg(T_{u \theta}+T_{\theta \phi}\frac{a}{r^2+a^2}\bigg)\dot{\Theta}\dot{U}-\frac{2 a\rho^2}{(r^2+a^2)^2} \bigg(T_{u \phi}+\frac{a}{r^2+a^2} T_{\phi\phi}\bigg)\dot{U}\dot{R}-2T_{\theta\phi}\frac{a \rho^2}{(r^2+a^2)^2} \dot{\Theta}\dot{R}+ T_{\phi \phi} \frac{a^2 \rho^4}{(r^2+ a^2)^4} \dot{R}^2.
\end{dmath}
To simplify the calculation, we consider only the leading order terms in $\Delta$ and such leading order terms are of the highest power in $\dot U$ as $\dot U \propto 1/\Delta$. So, in the leading order approximation, we have
 \begin{equation}
     \rho_0=\bigg(T_{u u}+ 2T_{u \phi}\frac{a}{r^2+a^2}+ T_{\phi\phi}\frac{a^2}{(r^2+a^2)^2}\bigg)\dot{U}^2.
 \end{equation}
Now, Substituting the values of $T_{\mu \nu}$ from Eqs.~\eqref{u46} - \eqref{u410} and the value of $\dot{U}$ from Eq.~\eqref{u436}, we get
\begin{equation}
    \rho_0=-\frac{4 \dot R^2 r \left(a^2 \sin ^2\theta M''+2 r M'\right)}{8 \pi \Delta^2}, \label{u439}
\end{equation}
resulting in the divergent energy density perceived by an infalling observer at $\Delta=0$. This implies that the firewall is not the artifact of spherical symmetry and does occur in axial symmetry also.

\section{Local pressure and flux}

\subsection{Kerr-Vaidya metric in advanced coordinates}

To calculate the pressure $p_0=T_{\mu\nu}n^\mu n^\nu$ perceived by an infalling observer, we need an outgoing spacelike normal $n^\mu$ satisfying $n^\mu u_\mu=0$ and $n^\mu n_\mu>0$. Such normal are not uniquely determined and one of them can be written as
\begin{equation}
    n_\mu=(-\dot R, \dot V,0,0).
\end{equation}
Using this normal, we now calculate the pressure for an infalling observer in the Kerr-Vaidya metric in advanced coordinates
\begin{equation}
    p_0=T_{\mu\nu}n^\mu n^\nu= T_{v v}{n^v}^2 +2 T_{v \phi}n^v n^\phi+ T_{\phi\phi}{n^\phi}^2.
\end{equation}
Again, direct calculation substituting the values of $T_{\mu \nu}$ from Eqs.~\eqref{v41} - \eqref{v45} and $\dot{V}$ from Eq.~\eqref{v427} gives
\begin{dmath}
p_0=\frac{M'}{16 \pi \dot R^2 \left(a^2+r^2\right)^2 \left(a^2 \cos 2\theta+a^2+2 r^2\right)^4} \bigg(2 a^6 \dot R^2+4 a^2 r^4 \left(4 a^2 \dot \Theta^2-3 \dot R^2+4\right)+8 \dot \Theta^2 r^8+4 r^6 \left(5 a^2 \dot \Theta^2+2\right)+a^4 r^2 \left(4 a^2 \dot \Theta^2-7 \dot R^2+8\right)+4 a^2 r^2 \cos 2\theta \left(a^4 \dot \Theta^2+2 a^2 \left(\dot R^2+\dot \Theta^2 r^2\right)+3 \dot R^2 r^2+\dot \Theta^2 r^4\right)-a^4 \dot R^2 \left(2 a^2+r^2\right) \cos 4\theta\bigg)\\ \bigg(-a^2 \dot R^2 \sin ^2\theta \left(a^2 \cos 2 \theta+a^2+2 r^2\right)+\left(a^2+ r^2\right)^2 \left(\dot \Theta^2 \left(a^2+2 r^2\right)+a^2 \dot \Theta^2 \cos 2\theta+ 2\right)\bigg).
\end{dmath}
As in the calculation of density, the terms containing $M''$ has been neglected during the calculation. Near the equatorial plane, $\dot{\Theta}=0$, and this expression for the pressure simplifies to
\begin{dmath}
p_0= -\frac{\left(a^4 \left(2 \dot R^2-1\right)+a^2 \left(3 \dot R^2-2\right) r^2-r^4\right) \left(a^4-a^2 \left(\dot R^2-2\right) r^2+r^4\right) M'}{16 \pi \dot R^2 r^6 \left(a^2+r^2\right)^2}.
\end{dmath}
We can similarly calculate the flux perceived by an infalling observer using the relation $\phi_0=T_{\mu\nu}u^\mu n^\nu$. Direct calculation near the equatorial plane $\dot \Theta=0$, after neglecting the terms containing $M''$ in the similar manner as above for pressure and density gives the following expression for flux
\begin{equation}
    \phi_0=\frac{\left(a^8-2 a^6 \left(\dot R^4-2\right) r^2-3 a^4 \left(\dot R^4-2\right) r^4+4 a^2 r^6+r^8\right) M'}{16 \pi \dot R^2 r^4 \left(a^2+r^2\right)^3}.
\end{equation}
As we are interested only in the qualitative analysis, we do not need the general expression for the flux which is more involved than density and pressure. Thus, like the local density, pressure and flux perceived by an infalling observer are not the artifact of spherical symmetry and does occur in axial symmetry also. Now, for $a=0$ we have
\begin{equation}
    \rho_0=p_0=\phi_0= \frac{M'}{16 \pi r^2 \dot R^2},
\end{equation}
which is the result for the local variables derived in Ref.~\cite{19} for spherical symmetry.

\subsection{Kerr-Vaidya metric in retarded coordinates}

Proceeding in a similar manner as for the advanced coordinates, we take the spacelike normal
\begin{equation}
    n_\mu=(-\dot R, \dot V,0,0),
\end{equation}
satisfying $n^\mu u_\mu=0$ and $n^\mu n_\mu>0$, but not uniquely determined. Using this normal, we can calculate the pressure for an infalling observer in Kerr-Vaidya metric in retarded coordinates as $p_0=T_{\mu\nu}n^\mu n^\nu$. Direct calculation substituting the values of $T_{\mu \nu}$ from Eqs.~\eqref{u46} - \eqref{u410} and the value of $\dot{U}$ from Eq.~\eqref{u436} gives
\begin{equation}
    p_0=-\frac{4 \dot R^2 r \left(a^2+r^2\right)^2 \left(a^2 \sin ^2\theta M''+2 r M'\right)}{8 \pi \rho^4 \Delta ^2},
\end{equation}
in the leading order Taylor's expansion. Like density, this expression for the pressure diverges as $\Delta \to 0$. Similarly, substitution of the values of $T_{\mu \nu}$ from Eqs.~\eqref{u46} - \eqref{u410} and the value of $\dot{U}$ from Eq.~\eqref{u436} onto the relation $\phi_0=T_{\mu\nu}u^\mu n^\nu$ gives the flux
\begin{equation}
    \phi_0=\frac{4 \dot R^2 r \left(a^2+r^2\right) \left(a^2 \sin ^2\theta M''+2 r M'\right)}{8 \pi \rho^2 \Delta ^2},
\end{equation}
in the leading order Taylor's expansion. This expression for the flux also diverges as $\Delta \to 0$. Now, we have, for $a=0$,
\begin{equation}
    \rho_0=p_0=-\phi_0=-\frac{8 \dot R^2 M'}{8 \pi (r-2 M)^2}.
\end{equation}

\section{Horizon crossing}

As mentioned in Sec.~\ref{114}, horizon crossing is possible if the distance $R(\tau)-r_g(T(\tau))\to 0$, and the horizon crossing does not happen if $\dot{R}-r_g'\dot{T}>0$ for some $R(\tau)-r_g(T(\tau))>0$. Here we take the Kerr-Vaidya metric in advanced coordinates and assume the quasi-stationary evaporation as was done in Sec.~\ref{114}. This assumption, as mentioned, enables us to neglect the backreaction on the metric because of evaporation. Under these settings
\begin{dmath}
\dot{R}-r_g'\dot{V}= \dot{R}-\bigg(-(\rho^2 \dot{\Theta^2}+1)\frac{r^2+a^2}{2 \rho^2 \dot{R}}+ \frac{a^2 \sin^2\theta}{2 (r^2+a^2)}\dot{R}\bigg)r_g'\\
= \bigg(1-\frac{a^2 \sin^2\theta}{2 (r^2+a^2)} r_g'\bigg)\dot{R}+(\rho^2 \dot{\Theta^2}+1)\frac{r^2+a^2}{2 \rho^2 \dot{R}}r_g',
\end{dmath}
where Eq.~\eqref{v427} has been used. Hence, horizon crossing by the test particle is not possible if
\begin{equation}
    |\dot{R}|<\sqrt{-\frac{(\rho^2 \dot{\Theta^2}+1) (r^2+a^2)^2 r_g'}{\rho^2(2(r^2+a^2)-a^2 \sin^2\theta r_g')}},
\end{equation}
note that, $r_g'<0$ here. Again, as mentioned in Sec.~\ref{114}, we can take the limit $r_g' \approx -\kappa/r_g^2$ for a quasi-stationary black-hole, $\kappa \sim 10^{-3}-10^{-4}$ being the surface gravity. This gives
\begin{equation}
    |\dot{R}|<\sqrt{-\frac{(\rho^2 \dot{\Theta^2}+1) (r^2+a^2) r_g'}{2 \rho^2}}.
\end{equation}
Near the equatorial plane, this expression simplifies to
\begin{equation}
    |\dot{R}|<\sqrt{-\frac{ (r^2+a^2) r_g'}{2 r^2}},
\end{equation}
and this gives the condition for which horizon crossing is prevented for a macroscopic rotating black-hole. Let us do a numerical estimation by assuming a slowly rotating macroscopic black-hole, such that, $a^2 \sim 0$ and $\sqrt{\kappa/r_g^2}\approx 0.01/r_g$. Then, the horizon crossing could happen only if $|\dot{R}|>0.005/r_g$.
This difficulty in crossing the horizon by a slowly-moving test particle is already pointed out in Sec.~\ref{114} for the case of spherical symmetry. Thus, there is no qualitative difference in this horizon crossing behavior of a test particle in either rotating or non-rotating black holes. Following the exact same lines of argument made in Sec.~\ref{114} for massless test particles, we can infer that massless particles always cross the apparent horizon of a rotating black hole.

\section{Quantum energy inequality}

It is mentioned in Sec.~\ref{113} that quantum energy inequality provides the bound on the allowed negative value of the energy density. For a field in arbitrary Hammard state $\omega$, the bound on the energy density $\rho_0$ as perceived by an observer moving in the trajectory $\xi$ is given by~\cite{23}
\begin{equation}
    \int_\xi {d\tau \mathscr{F}^2(\tau)\rho(\tau)} \geqslant {\cal B}_A(R, \mathscr{F}, \xi) \label{qe456},
\end{equation}
where $\mathscr{F}$ is a real valued function with compact support, $R$ is the Ricci scalar and ${\cal B}_A \geqslant 0$. 
The calculation here proceeds in the similar way as in the case of the spherical symmetry given in Sec.~\ref{113}. We first consider an advancing horizon $r_g'(u)>0$. We then choose the trajectory $\xi$ of an infalling observer such that the energy density given by Eq.~\eqref{u439} is valid. As already pointed out, we are considering the semiclassical limit, which assumes the curvature near the horizon small enough such that Eq.~\eqref{qe456} is applicable. Also, in this limit, the horizon radius and the mass of a black hole does not change appreciably as an observer approaches towards its vicinity. Hence, $\dot M=dM/d\tau=M'(U)\dot U\approx \textrm{constant}$. Given the trajectory $\xi$, we can choose  $\mathscr{F}\approx 1$ at the horizon crossing and $\mathscr{F}\to 0$ within the NEC-violating domain. So, as the trajectory $\xi$ approaches the horizon, the left hand side of Eq.~\eqref{qe456} takes the form
\begin{align}
    \int_\xi {d\tau \mathscr{F}^2(\tau)\rho(\tau)} & =-\int_\xi {\frac{4 \dot R^2 r_g \left(a^2 \sin ^2\theta M''+2 r_g M'\right)}{8 \pi \Delta^2} d\tau}\\
    & \approx \int_\xi {\frac{8 r_g^2 \dot M}{16 \pi \Delta (r_g^2+a^2)} dR} \propto \log \Delta \to -\infty,
\end{align}
where $\rho_0$ and $\dot U$ has been substituted from Eq.~\eqref{u439} and Eq.~\eqref{u436} respectively for simplification. Similarly, as in some calculations above,we have neglected the terms containing $M''$ as $M''\ll M'/M$ holds in the semiclassical limit. This shows that the quantum energy inequality is violated in some region close to the horizon of the Kerr-Vaidya black hole in advanced coordinates. This violation of the quantum energy inequality forbids the growth of the black holes once they are formed. Otherwise, the semiclassical approximation is not valid near the horizon.

\let\cleardoublepage\clearpage

\begin{savequote}[45mm]
Death is as sure for that which is born, as birth is for that which is dead. Therefore grieve not for what is inevitable.
\qauthor{Gita: Chapter 2, Verse 27}
\end{savequote}

\chapter{Conclusion and discussions}

In this project, we have worked on near horizon properties of the Kerr-Vaidya metrics. Kerr-Vaidya metrics being the simplest non-stationary generalization of the Kerr metric. They can be derived from the Vaidya metrics by the complex Newman-Janis coordinate transformation.

Only the Kerr-Vaidya metric in retarded coordinates was derived from the Vaidya metric in retarded coordinates using the complex coordinate transformation of Newman and Janis. Following their procedure, we have here developed the Newman-Janis like coordinate transformation and obtained the axisymmetric Kerr-Vaidya metric in the advanced coordinates.

We have also located the position of an apparent horizon for the Kerr-Vaidya metrics, both in advanced and retarded coordinates. We have verified that the apparent horizon of the Kerr-Vaidya metric in advanced coordinates coincides with the event horizon $r_0=M+\sqrt{M^2-a^2}$ of the Kerr metric. This result has been obtained already by various authors in the past~\cite{nj3}. However, the location of the apparent horizon for the Kerr-Vaidya metric in advanced coordinates differs from the event horizon of the Kerr metric by a factor proportional to $M_v'$, that is, $r_g-r_0\propto M_v'$ for the ingoing Kerr-Vaidya metric. This factor is small $|M_v'|\ll 1$ in our semiclassical domain. However, we have found that the commonly used assumption $z'(0)=0$ does not hold at the poles. But the condition $z(0)=0=z(\pi)$ does hold necessarily to preserve the symmetry of the problem and to reduce the solution to the spherical symmetry in the particular case of $a=0$.

There is also another remarkable feature that the Kerr-Vaidya geometries show. While these space-times violate the null energy condition for all $a\ne 0$, their energy-momentum tensors belong to the classically impossible type \Romannum{3} form of the Hawking-Ellis/Segre classification. The violation of the null energy condition, which could have been considered as a drawback in the classical domain, is a necessary consequence of both the emission of the Hawking radiation and the finite time formation of the trapped surface according to a distant observer. The results are quite different for the special case of $a=0$: the null energy condition is violated for $M_v'<0$ and $M_u'>0$ and the energy-momentum tensor, in both cases, reduce to the standard type \Romannum{2} form of the Hawking-Ellis classification provided the energy condition holds.

Calculations of local variables: energy density, pressure, and flux in the frame of an infalling observer give no astounding results in advanced coordinates. They, being finite quantities reduces to the case where all three of them are equal when $a=0$, a result already obtained for spherical symmetry~\cite{18}. We have also found that the horizon crossing happens in a finite time in the clock of a distant observer, unlike the case in a classical eternal black hole. Calculations of local variables, however, give some fascinating results in retarded coordinates. As in advancing spherically symmetric apparent horizon, the outgoing Kerr-Vaidya metric leads to a firewall: the divergent value of energy density, pressure, and flux perceived by an infalling observer. This demonstrates that the firewall is not an artifact of spherical symmetry. Moreover, this firewall leads to the violation of quantum energy inequality, thereby implying that the apparent horizon, once formed, cannot grow.

Our assumption of $a=\mathrm{constant}$ is not compatible with the black holes evaporating eventually. One of the reasons is for $M<a$, the real root of $\Delta=0$ does not exists, and this implies the possibility of the existence of a naked singularity. Another reason is that the Hawking temperature~\cite{ht68}
\begin{equation}
    T=\frac{1}{2\pi}\frac{r_0 -M} {r_0^2+a_0^2},
\end{equation}
approaches zero as $M\to a$. Furthermore, the study done by Arbey \etal~\cite{va69} shows that $a/M$ decreases faster than $M$ during evaporation. However, this variation in $a$ should not affect the result of the firewall in retarded coordinates. The reason behind this is the firewall is manifested as a result of $\Delta=0$, immediately after the horizon formation, and occurs even in spherical symmetry $a=0$.

In the near future, we intend to extend the self-consistent approach to the more general axisymmetric metric (obviously, first relaxing the assumption of $a=\mathrm{constant}$). The obvious extension of our works includes the application of the self-consistent approach to other metric theories of gravity. This extension for the spherically symmetric geometry is done in Ref.~\cite{st70}.

\let\cleardoublepage\clearpage

\appendix

\chapter{Apparent horizon of Kerr-Vaidya metric in advanced coordinates} \label{a1}

After setting $l^v=1$ by utilizing the freedom of normalization in Eq.~\eqref{h34}, the leading order components of the future-directed outward-pointing null vector orthogonal to the two-surface $r=r_0+z(\theta)$ are
\begin{align}
    l^v &=1, \qquad l^r=\frac{(r_0^2-a^2)z'}{2r_0(r_0^2+a^2)},\qquad
    l^\theta =  -\frac{z'}{r_0^2+a^2}, \\
    l^\phi &=\frac{a}{r_0^2+a^2}+\frac{a(a^4-7a^2 r_0^2-10r_0^4-a^2(r_0^2-a^2)\cos2\theta)z}{4r_0(r_0^2+a^2)},
\end{align}
where we assume that $z\ll r_0$ and $z'\ll r_0$. Similarly, the derivatives of these future-directed outward-pointing null vector's components the leading order are
\begin{align}
    \frac{dl^r}{d\theta}=\frac{(r_0-M) z'(\theta)}{2 r_0 M}, \qquad \frac{dl^\theta}{d\theta}=-\frac{z''(\theta)}{2 r_0 M}.
\end{align}

Now, to locate an apparent horizon using Eq.~\eqref{kvh311}, we need to know the explicit values of $\sigma_P$ and $\sigma_Q$. To calculate $\sigma_P$, we use the relation $[\sigma_{i j}]_P=e^a_{i P} [h_{a b}]_P e^b_{j P}$, where $[h_{a b}]_P$ is given by Eq.~\eqref{46} is calculated at $r=r_0+z(\theta)$. Similarly, to calculate $\sigma_Q$, we use the relation $[\sigma_{i j}]_Q=e^a_{i Q} [h_{a b}]_Q e^b_{j Q}$, where $[h_{a b}]_Q$ can be calculated using the relation
\begin{equation}
    \begin{split}
        [h_{a b}]_Q= &[h_{a b}]_P+ \delta [h_{a b}]_P\\
        =& [h_{a b}]_P+ \left(l^2 \partial_r +l^3 \partial_\theta +M_v \partial_v\right)[h_{a b}]_P dv.
    \end{split}
\end{equation}
Here, $l^2$, $l^3$ and their derivatives are calculated from Eq.~\eqref{h34}.
Now, we substitute the relations for $\sigma_Q$ and $\sigma_P$ in Eq.~\ref{kvh311} to write the equation of the apparent horizon explicitly up to the first order in Taylor's expansion
\begin{equation}
    l^r h_{33} \frac{dh_{22}}{dr}+ l^r h_{22} \frac{dh_{33}}{dr} +h_{33} l^\theta \frac{dh_{22}}{d\theta} +h_{22} l^\theta \frac{dh_{33}}{d\theta} +h_{22} \frac{dh_{33}}{dv} +2 h_{22} h_{33} \frac{dl^\theta} {d\theta}=0.
    \label{a4}
\end{equation}
Here, the metric $h_{ab}$ and its derivatives are calculated at some arbitrary point $P$ on the two surface $r=r_0+z(\theta)$ and $z\ll r_0$ and $z'\ll r_0$ is assumed. We can calculate the values of $h_{ab}$ and its derivatives directly from Eq.~\eqref{46}. We then substitute the values of $h_{ab}$, $l^\mu$ and their derivatives in Eq.~\eqref{a4} to get Eq.~\eqref{kvh313}.

An alternative derivation is based on extending the vector field $l^\mu$ from the hypersurface  $v=\text{constant}$ to the bulk in such a way that the new field $\breve{l}^\mu$ satisfies the geodesic equation $\breve{l}^\mu_{;\nu}\breve{l}^\nu=0$. In fact, this needs to be done only on the hypersurface itself, where it is realized by setting
\begin{equation}
    \breve{l}^\mu=l^\mu, \qquad \breve{l}^\mu_{;m}=l^\mu_{\,;m}, \qquad   \breve{l}^\mu_{\,;0}=-l^\mu_{\,;m} l^m,
    \label{a6}
\end{equation}
for $m=1,2,3$, and $\breve{l}^0=l^0=1$.

For the affinely parameterized geodesic congruence $\vartheta_l=\breve{l}^\mu_{;\mu}$, the substitution from the Eq.~\eqref{a6} gives
 \begin{equation}
     \vartheta_l=-l^0_{\,;m} l^m+l^m_{;m}=0.
 \end{equation}
Direct calculation from this equation leads to  Eq.~\eqref{kvh313}. To solve Eq.~\eqref{kvh313} numerically, we normalize all the quantities occurring in the equation in the unit of $M$ by substituting $M'\vcentcolon=-\kappa/M^2$, $a/M\vcentcolon=\alpha$,
$x\vcentcolon= r_0/M$ and then set $M\to 1$ to get:
\begin{align}
    &\left( \sqrt{1-\alpha^2}\big(4 x -
    \sqrt{1-\alpha^2}\,(1 -\sqrt{1-\alpha^2}\,) \sin^2\theta\big)-\frac{  4 \alpha^2x^2\kappa\sin^2\theta  }{x^2+\alpha^2\cos^2\theta} \right)z
   +4x(\cot\theta z'+z'') \nonumber\\
   &-2 \alpha^2x\kappa\sin^2\theta =0.
\end{align}

\chapter{Energy-momentum tensor of the Kerr-Vaidya metric}\label{a2}

The non-zero components of the energy-momentum tensor for the Kerr-Vaidya metric in advanced coordinates are
\begin{align}
    &  T_{v v}=\frac{ r^2(a^2+ r^2)- a^4 \cos^2\theta\sin^2\theta}{4\pi\rho^6}M'
        -\frac{a^2 r \sin^2\theta} {8\pi\rho^4} M'' \label{v41},\\
    & T_{v \theta}=-\frac{ a^2 r \sin\theta \cos\theta}{4\pi \rho^4}M',\\
    & T_{v \phi} = - a \sin^2\theta T_{v v}-a \sin^2\theta\frac{r^2-a^2\cos^2\theta}{8\pi\rho^4}M', \\
    & T_{\theta \phi}=\frac{ a^3 r \sin^3\theta \cos\theta}{4\pi \rho^4}M',\\
    & T_{\phi \phi}= a^2 \sin^4\theta T_{v v}+ a^2 \sin^4\theta\frac{r^2-a^2\cos^2\theta}{4\pi\rho^4}M'. \label{v45}
\end{align}
Similarly, the non-zero components of the energy-momentum tensor for the Kerr-Vaidya metric in retarded coordinates are
\begin{align}
    & T_{u u}=-\frac{r^2( a^2 +r^2)- a^4 \cos^2\sin^2\theta}{4\pi\rho^6}M'
    -\frac{a^2 r \sin^2\theta}{8\pi\rho^4}M'', \label{u46}\\
    & T_{u \theta}=-\frac{2 a^2 r \sin\theta \cos\theta}{8\pi \rho^4}M',\\
    &  T_{u \phi}=-  a\sin^2\theta T_{u u}+a\sin^2\theta \frac{r^2-a^2\cos^2\theta}{8 \pi\rho^4}M',\\
    & T_{\theta \phi}=\frac{2 a^3 r \sin^3\theta \cos\theta}{8\pi \rho^4}M',\\
    &  T_{\phi \phi}= a^2 \sin^4\theta T_{u u}-a^2 \sin^4\theta\frac{(r^2-a^2\cos^2\theta)}{4\pi\rho^4}M'.
    \label{u410}
\end{align}

The orthonormal tetrad with where $w^\mu=e_{\hat 1}^\mu+e_{\hat 0}^\mu$ is given by:
\begin{align}
    &e_{\hat 0}= \left(-1,\pm rM/\rho^2,0,0\right),    \\
    &e_{\hat 1}= \left(1,\pm\left(1-rM/\rho^2\right),0,0\right),  \\
    &e_{\hat 2}=\left(0,0,1/\rho,0\right), \\
    &e_{\hat 3}=\frac{1}{\rho}\left(a \sin \theta,\pm a \sin\theta,0,\csc\theta\right),
\end{align}
where, $+/-$ sign is for advanced/retarded coordinates. These tetrads satisfiy the following orthonormality and completeness relations:
\begin{align}
    & e_{\hat a}^\mu e_{\hat b \mu}=0 \qquad \textrm{for} \qquad \hat a \ne \hat b\\
    & e_{\hat 0}^\mu e_{\hat 0 \mu}=-1\\
    &e_{\hat a}^\mu e_{\hat a \mu}=1 \qquad \textrm{for} \qquad \hat a \ne 0.
\end{align}

\backmatter

\chapter{List of Symbols}


We present the following list of the symbols and conventions used in the text:
\begin{list}{}{%
\setlength{\labelwidth}{24mm}
\setlength{\leftmargin}{0mm}}
\item We use the metric signature $-+++$.
\item We use the Planck units, $\hbar=c=G=1$.
\item Covariant derivative is denoted by the semicolon ($;$).
\item $g_{\mu\nu}$ is the metric of $1+3$-dimensional space-time and $g^{\mu\nu}$ is its contravariant counterpart.
\item $\tau$ denotes the proper time.
\item $i$ denotes $\sqrt{-1}$.
\item $\delta^\mu_\nu$ denotes the Kronecker delta.
\item Except where mentioned $x'(t)$ denotes the derivative of $x$ with respect to $t$ and $\dot x(t)$ denotes the derivative of $x$ with respect to $\tau$.
\item $r_g$ represents an apparent horizon.
\item $r_0$ is the solution of the equation $r^2-2 M r +a^2=0$.
\item $\rho^2=r^2+a^2 \cos^2\theta$.
\item $\Delta=r^2+a^2-2 M r$.
\item $R_{\mu\nu}$ denotes the Ricci tensor and $R$ denotes the Ricci Scalar.
\item Einstein tensor $G_{\mu\nu}=R_{\mu\nu}- \frac{1}{2} g_{\mu\nu}R$.
\item $T_{\mu\nu}$ denotes the energy-momentum tensor.
\item $\kappa$ denotes the surface gravity.
\item Lie derivative of the vector $A^\alpha$ along the curve $\xi$ is defined as $\textit{\pounds} =A^\alpha_{;\beta}l^\beta-l^\alpha_{;\beta}A^\beta$ where $l^\alpha=\frac{dx^\alpha}{d\lambda}$ is the tangent vector to the curve.
\end{list}

\let\cleardoublepage\clearpage

\bibliography{thesis}

\end{document}